\documentclass[english,aps,prl,amsmath,amssymb,twocolumn,letterpaper,showpacs,superscriptaddress]{revtex4-1}
\usepackage{babel}
\usepackage{verbatim}
\usepackage{amsmath}
\usepackage{amssymb}
\usepackage{graphicx}
\usepackage{color}
\usepackage[unicode=true,pdfusetitle,
 bookmarks=true,bookmarksnumbered=false,bookmarksopen=false,
 breaklinks=false,pdfborder={0 0 1},backref=false,colorlinks=false]
 {hyperref}
\usepackage{pdfpages}	
\usepackage{pgffor}		

\makeatletter
\AtBeginDocument{\let\LS@rot\@undefined} 
\makeatother							 

\makeatother

\begin{document}

\title{Noise-induced backscattering in a quantum-spin-Hall edge}

\author{Jukka I. V\"ayrynen}

\affiliation{Station Q, Microsoft Research, Santa Barbara, California 93106-6105,
USA}

\author{Dmitry I. Pikulin}

\affiliation{Station Q, Microsoft Research, Santa Barbara, California 93106-6105,
USA}

\author{Jason Alicea}

\affiliation{Department of Physics and Institute for Quantum Information and Matter,
California Institute of Technology, Pasadena, CA 91125, USA}

\affiliation{Walter Burke Institute for Theoretical Physics, California Institute
of Technology, Pasadena, CA 91125, USA}

\date{\today}
\begin{abstract}
Time-reversal symmetry suppresses electron backscattering in a quantum-spin-Hall edge,
yielding quantized conductance at zero temperature. Understanding
the dominant corrections in finite-temperature experiments
remains an unsettled issue. We study a novel mechanism for conductance
suppression: backscattering caused by incoherent electromagnetic noise.
Specifically, we show that an electric potential fluctuating randomly in time can backscatter
electrons inelastically without constraints faced by electron-electron
interactions.  We quantify noise-induced corrections to the dc
conductance in various regimes and propose an experiment to test this scenario.
\end{abstract}

\maketitle

\textbf{\emph{Introduction.}}~From a technological perspective, 
the main promise of two-dimensional topological insulators (2D TIs)
stems from their edge states, which are protected by a combination of symmetry and topology~\cite{2005PhRvL..95n6802K,2005PhRvL..95v6801K,2006PhRvL..96j6802B}.
The `helical' low-energy edge spectrum consists of \emph{degenerate} counterpropagating electron states with opposite spins as required by time-reversal symmetry. Kramers orthogonality of the two states prevents elastic backscattering by
a static potential, in turn yielding a quantized zero-temperature conductance 
$G = G_{0}\equiv e^{2}/h$ per edge.  Perfect quantization has, however,
so far eluded experimental observation~\cite{2007Sci...318..766K,2009Sci...325..294R,2014PhRvB..89l5305G,Knez14,Du2015,1367-2630-18-8-083005,PhysRevLett.119.056803,2017NatPh..13..677F,Wu76}.

\begin{figure}
\includegraphics[width=1\columnwidth]{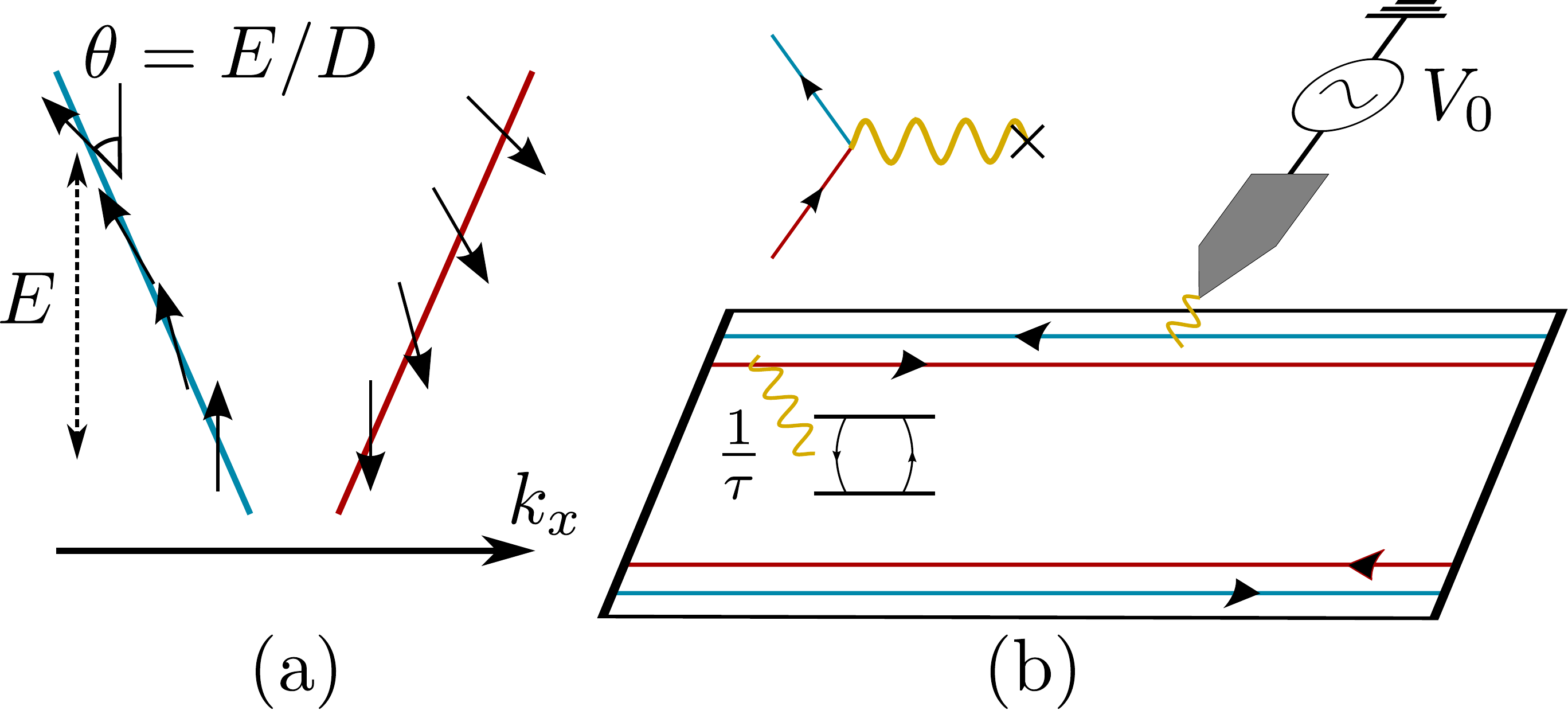}\caption{(a) Spin texture in momentum space. Over a small energy interval $E$,
the spin of an eigenstate rotates by a small angle $E/D$. (b) 
A time-dependent scalar potential 
induces backscattering when such spin textures exist.  
The potential can arise from a fluctuating two-level system near the edge or from an ac voltage $\sim \cos\omega_0t$ applied by a nearby gate; Eqs.~(\ref{eq:singleTLSfinal}) and (\ref{eq:deltaGexp}) predict the respective decrease in two-terminal dc conductance arising from these sources.  Note that an applied ac voltage provides a controlled way of probing the spin texture required in our scenario.
\label{fig:1}}
\end{figure}

In practice, it was realized early on that many \emph{inelastic} 
effects circumvent band-topology constraints and can hinder
the edge-mode propagation by introducing backscattering~\cite{2006PhRvB..73d5322X,2006PhRvL..96j6401W,2009PhRvL.102y6803M,2011PhRvL.106w6402T,2012PhRvL.108h6602B,schmidt_inelastic_2012,2013PhRvL.110t6803C,2013PhRvL.110u6402V,2013PhRvL.111h6401A,2014PhRvB..90k5309V,2014PhRvB..90g5118K,2015PhRvL.115r6404C}.
These backscattering mechanisms reflect the fact that time-reversal
symmetry allows \emph{non-degenerate} counterpropagating states to
have overlapping spin wave functions.  Such non-zero overlap occurs generically  in systems with fully broken spin-rotation symmetry. 
Indeed, it is well known that structural or bulk inversion asymmetry
 in 2D TIs may induce nontrivial edge spin texture in momentum space~\cite{schmidt_inelastic_2012}; that is,
the edge-state spin quantization axis ``rotates'' as a function
of momentum~\footnote{The spin quantization can also be broken in a disordered manner; in
this paper we ignore these terms for simplicity.  Note that they do
not lead to significant backscattering at low temperature~\cite{PhysRevLett.116.086603}. 
Electron-electron interactions together with Rashba spin-orbit disorder lead to a correction to conductance that vanishes as $T^4$ (or with a higher power law) as $T \to 0$~\cite{PhysRevB.96.155134}.} as sketched in Fig.~\ref{fig:1}a. The necessary energy transfer for backscattering
was considered to originate from thermal itinerant edge electrons or phonons,
or fluctuating localized spins.  Apart from the latter, the backscattering
rate was found to be strongly suppressed at low temperatures:  Electron-electron
interactions perturbatively produce a conductance correction $\delta G\equiv G_{0}-G\propto T^{4}$
at low temperatures~\cite{PhysRevB.85.235304,schmidt_inelastic_2012,2014PhRvB..90g5118K}, while phonon scattering is even more suppressed.
Localized spins~\footnote{Localized spins may occur for example due to charge puddles~\cite{2013PhRvL.110u6402V,2014PhRvB..90k5309V} or nuclear spins~\cite{Hsu2017,*Hsu2018} } impart a stronger effect in the perturbative limit, $\delta G\propto\ln^{2}T$, but their presence 
 need not be a universal feature of all 2D TI materials. 

In this paper we show that a time-dependent scalar potential  might
dominate the backscattering in practice. This mechanism is expected
to be ever-present in all materials in the form of electrical noise
and is almost free from phase-space constraints. 

We start from a qualitative derivation of our main result, the estimate
of the decrease $\delta G$ in the edge dc conductance.  
Spin texture in momentum space makes the  edge electron density operator off-diagonal in the basis of right and left movers~\cite{PhysRevLett.116.086603}.
%
Coupling the total density to a scalar potential
$U(x)$ thus produces an effective backscattering matrix element
 $V_{pR\to p'L}$ that, for small energy transfer $v|p+p'|$, vanishes
as $V_{pR\to p'L}=\frac{v}{D}(p+p')U_{2k_{F}}$. 
(We use $\hbar=k_B =1$ units.)
Here $p,\,p'$ are momenta measured from the
Fermi points $\pm k_{F}$ and $U_{2k_{F}}$ is the $2k_{F}$ Fourier
component of the potential; $v$ is the edge mode velocity and $D/v$ is the momentum scale  over which the spin rotates; see Fig.~\ref{fig:1}a. 
For a potential $U$ fluctuating harmonically 
with frequency $\omega$, the backscattering rate is $\Gamma=2\pi\nu\frac{\omega^{2}}{D^{2}}|U_{2k_{F}}|^{2},$
where $\nu=1/2\pi v$ is the edge density of states per length. When
a bias voltage $V$ is applied across the edge, $\nu eV$
states contribute to the current. The backscattered
current at low temperature thus reads $\delta I_{\omega}=2\pi e^{2}\nu^{2}\frac{\omega^{2}}{D^{2}}|U_{2k_{F}}|^{2}V $.
One needs to integrate $\delta I_{\omega}$ over the full noise spectrum.
A thermal noise source at temperature $T$  can only emit photons of frequencies $\omega \lesssim T$, which cuts off the integration over $\omega$. 
For low-frequency ($\omega \ll T$) noise caused by a single fluctuating electric dipole, modeled as a two-level system (TLS) with relaxation rate $\tau^{-1} \ll T$, the noise spectrum is Lorentzian and 
the integration yields
\begin{equation}
\delta G\sim   G_0 \frac{T |U_{2k_{F}}|^{2}}{  D^2 \tau v^2}\,,\label{eq:singeTLSqualit}
\end{equation}
which is one of our main results. The refined version of this equation
appears in Eq.~(\ref{eq:singleTLSfinal}). On a long edge, many
dipoles contribute incoherently to $\delta G$, leading to resistive
edge transport. The long-edge resistance is obtained by summing Eq.~(\ref{eq:singeTLSqualit})
over impurities and averaging over $\tau$. Assuming a distribution of relaxation times
$P(\tau)\sim1/\tau$  and a short-time cutoff $\tau_{0}$,
the resistance becomes 
\begin{equation}
R\sim Ln G_{0}^{-1} \frac{T |U_{2k_{F}}|^{2}}{D^2\tau_{0}v^{2}}\,,\label{eq:ResistanceQualit}
\end{equation}
with $L$ the length of the edge and $n$ the number of dipoles per length.  
Equations~(\ref{eq:singeTLSqualit}) and (\ref{eq:ResistanceQualit})
are valid at `high' temperatures where the dipoles are not frozen. In
this regime the mechanism does not lead to strong temperature dependence
of $R$, unlike conventional backscattering processes arising for example from electron-electron interactions on the edge. 

\textbf{\emph{Model and derivation.}}~The Hamiltonian of a clean
helical edge is~\cite{schmidt_inelastic_2012} 
\begin{equation}
H_{0}=\int\frac{dk}{2\pi}(\varepsilon_{k}c_{kR}^{\dagger}c_{kR}+\varepsilon_{-k}c_{kL}^{\dagger}c_{kL})\,,\label{eq:H0}
\end{equation}
with $\varepsilon_{k}=vk-\mu$ the spectrum linearized about the
chemical potential $\mu=vk_{F}$.  We stress that $H_{0}$ does not assume
spin conservation and allows for a spin texture in momentum space.
The spin of an eigenstate follows from the unitary transformation
\begin{equation}
\left(\begin{array}{c}
c_{k\uparrow}\\
c_{k\downarrow}
\end{array}\right)=B_{k}\left(\begin{array}{c}
c_{kR}\\
c_{kL}
\end{array}\right)\, \label{eq:texture}
\end{equation}
that relates fermions with spin $\uparrow,\downarrow$ to left and right movers. Unitarity
and time-reversal symmetry impose $B_{k}^{\dagger}=B_{k}^{-1}$ and
$B_{k}=B_{-k}$. 

Consider next a time-dependent scalar potential that couples to the edge electron
density $\rho=\sum_{\sigma=\uparrow,\downarrow}\psi_{\sigma}^{\dagger}\psi_{\sigma}$:
\begin{equation}
H_{U}(t)=\int dx\rho(x)U(x)w(t)\,.\label{eq:Hv}
\end{equation}
We assume here that the noise-induced potential has separable dependence
on position and time, parametrized by $U(x)$ and $w(t)$, respectively.
This assumption certainly holds for telegraph noise (two-level-system
noise) from a single impurity, which we consider later. 
In the presence of $H_U(t)$, time-reversal symmetry is clearly broken, but is maintained in a time-averaged sense.

Using Eq.~(\ref{eq:texture}) to express the density $\rho$
in the $L/R$-basis, we see that $H_{U}$ does not conserve the number
of left and right movers.  In momentum space, the off-diagonal
part of Eq.~(\ref{eq:Hv}) reads 
\begin{equation}
H_{U,RL}(t)=w(t)\int\frac{dk}{2\pi}\frac{dk'}{2\pi}[B_{k'}^{\dagger}B_{k}]_{10}U_{k-k'}c_{k'R}^{\dagger}c_{kL}+h.c.\,,\label{eq:Hvwithtexture}
\end{equation}
where $[M]_{10}$ denotes the off-diagonal component of the 2x2 matrix $M$. 
 Equation~(\ref{eq:Hvwithtexture}) gives rise to a non-zero backscattering
current operator $\delta I(t)=-\frac{1}{2}e{d(N_{R}-N_{L})/dt}$.
We evaluate the average backscattering current $\left\langle \delta I(t)\right\rangle $
using the Kubo formula~\cite{kubo1957statistical}, treating $H_{U,RL}$ as a time-dependent
perturbation. We find 
\begin{flalign}
\left\langle \delta I(t)\right\rangle  & =e\int\frac{dk}{2\pi}\frac{dk'}{2\pi}|[B_{k'}^{\dagger}B_{k}]_{10}|^{2}|U_{k-k'}|^{2}(f_{-kL}-f_{kR})\label{eq:currentgeneralfinal}\\
 & \times2\text{Re}\int_{-\infty}^{0}dt'e^{-i(vk+vk'+i0)t'}w(t)w(t'+t)\,. \nonumber
\end{flalign}
Here we introduced Fermi functions
$f_{k\alpha}=f(\alpha vk-\mu_{\alpha})=\langle c_{k\alpha}^{\dagger}c_{k\alpha}\rangle$
with $f(E)=(e^{E/T}+1)^{-1}$; we identify here $R\equiv +$ and $L\equiv -$. 
Treating backscattering as a weak perturbation, we use  unperturbed Fermi functions where the bias voltage $V$ is incorporated by setting   chemical potentials $\mu_{R,L}=\mu\pm\frac{1}{2}eV$ for right and left movers.
As a sanity check, static perturbations with $w(t)=$ constant impose $k=-k'$ and thus $[B_{k'}^{\dagger}B_{k}]_{10} =0$ in Eq.~(\ref{eq:currentgeneralfinal}), implying that backscattering does not arise.
We will next take the linear-response limit
$eV\ll T$, where $f_{-kL}-f_{kR}\approx f(vk-\mu)[1-f(vk-\mu)]eV/T$. 

The time-averaged backscattered current, 
$\overline{\left\langle \delta I\right\rangle }=\mathcal{T}^{-1}\int_{0}^{\mathcal{T}}dt\left\langle \delta I(t)\right\rangle $
with $\mathcal{T}\to\infty$, is determined by the Fourier transform
of the correlator $\overline{w(t)w(t'+t)}$.
In terms of the power spectral density $S(\omega)=\int_{-\infty}^{\infty}dt'e^{i\omega t'}\overline{w(t)w(t'+t)}$,
the correction to the dc conductance $\delta G=d\overline{\left\langle \delta I\right\rangle }/dV$
can be written as 
\begin{flalign}
\delta G & =e^{2}\int\frac{dk}{2\pi}\frac{dk'}{2\pi}|[B_{k'+k_{F}}^{\dagger}B_{k+k_{F}}]_{10}|^{2}\label{eq:deltaGgeneral}\\
 & \times|U_{k+k'+2k_{F}}|^{2}T^{-1}f(vk)[1-f(vk)]S(vk-vk')\,.\nonumber 
\end{flalign}
Since $S(\omega)$ is non-negative, the
noise always decreases the dc edge conductance, i.e., $\delta G>0$.

The integrals in Eq.~(\ref{eq:deltaGgeneral}) cannot be explicitly
evaluated without specifying the function $B_{k}$ that encodes the spin texture. 
However, tractable results can be obtained when the vector $\mathbf{d}_{k}$ specifying the texture via $B_{k} = \exp {i\mathbf{d}_{k}\cdot\boldsymbol{\sigma}}$ has a slowly varying magnitude ${d}_{k}$ and a fixed direction, $\mathbf{d}_{k} = {d}_{k} \hat{\mathbf{n}}$. 
The former assumption can be used in Eq.~(\ref{eq:deltaGgeneral}) to expand $[B_{k'+k_{F}}^{\dagger}B_{k+k_{F}}]_{10} \approx n_{\perp} v{(k-k')}/ D_{k+k_F} $ with $n_{\perp} = [\hat{\mathbf{n}} \cdot\boldsymbol{\sigma}]_{10}$. 
Here $D_{k+k_F} = v/ \partial_k {d}_{k+k_F}$ defines the typical energy scale of the spin rotation (see Fig.~\ref{fig:1}).  
The expansion is valid when the noise spectrum is peaked at low frequencies so that $v(k-k') \ll D_{k+k_F}$ in the relevant region of integration in Eq.~(\ref{eq:deltaGgeneral}).  
Further assuming  $v{(k-k')},~T \ll \mu$, we obtain 
\begin{equation}
\delta G  \approx \frac{G_0}{v^2} |U_{2k_{F}}|^{2}  n_{\perp}^2 D^{-2} \int \frac{d \omega}{2\pi} \omega^2 S(\omega)  \,, \label{eq:deltaGgeneralintermediate}
\end{equation}
where  $D=D_{k_F}$. 
Equation~(\ref{eq:deltaGgeneralintermediate}) is  valid for any $\mu/D$. 
However, close to the Dirac point, $k_F =0$, the spin texture becomes quadratic (because of the property $d_{k}=d_{-k}$ following from time-reversal symmetry) with a curvature $k_0^{-1}$. 
When, $\text{max}(\mu,\,T) \ll v k_0$, one can replace $D^{-1}\to \text{max}(\mu,\,T) / (vk_0^2)$ in Eq.~(\ref{eq:deltaGgeneralintermediate}).

\textbf{\emph{Telegraph noise from a charge puddle.}}~The
conductance correction $\delta G$ in Eq.~(\ref{eq:deltaGgeneralintermediate})
depends on the noise spectrum. A realistic source is telegraph noise
caused by a charge puddle~\cite{2013PhRvL.110u6402V} with fluctuating
charge. 
We model the puddle as a quantum dot that creates a local edge electric potential $W(x,t)$ dependent on the dot's configuration at a given time $t$.
We assume that the dot has a sizable charging energy $E_{C}\gtrsim T$ so that only two charge states need to be retained.  To simplify our description we will further ignore different states of the puddle within the same charge sector, which is justified if noise predominantly arises from electric dipole fluctuations contributed by different charge states.  
In this case, the puddle acts as a two-level system  akin to a fluctuating
dipole,~\footnote{See Supplementary Material, where we consider noise from fluctuating dipoles in the dielectric as well as a quantum mechanical treatment of $w(t)$.}
 and its potential admits the separable form $W(x,t) = U(x)w(t)$ employed in Eq.~(\ref{eq:Hv}).
Here $U(x)$ is the effective dipole potential (the difference in
the potential in the two charge states) and $w(t)$ represents telegraph noise.  

The charge puddle's classical noise spectrum is given by~\cite{9780511551666}
\begin{equation}
S(\omega)=p_{0}(1-p_{0})\frac{2\tau}{1+\omega^{2}\tau^{2}}\,,\quad (\omega \ll T)\,,\label{eq:TLSNoise}
\end{equation}
where $\tau^{-1}$ is the relaxation rate of the excited state and
$p_{0}$ is the probability for the TLS to be in its ground state.
For a thermal population we have $p_{0}=1/(1+e^{-\Delta/T})$ with $\Delta=2E_{C}|\{N_{g}\}-\frac{1}{2}|$
the energy difference between the puddle's excited and ground states; $\{\dots\}$ denotes the positive fractional part while $N_{g}$ is 
a dimensionless parameter  determined by the puddle's electrostatic
environment (\textit{e.g.}, a neighboring puddle) and thus varies between different TLSs.
We treat $N_{g}$
as a uniformly distributed random variable.  Note, however,
that $N_{g}$ depends linearly on the edge chemical potential
$\mu$, which is tunable by a global gate. Therefore, due to the factor
$p_{0}(1-p_{0})$ in Eq.~(\ref{eq:TLSNoise}), we expect to see temperature-broadened
resonances in $\delta G$ of a short edge as the gate voltage is tuned;
see Eq.~(\ref{eq:singleTLSfinal}) below. Gate-induced conductance fluctuations 
are consistent with experiments in 
existing 2D TI candidate materials~\cite{2007Sci...318..766K,Li15,2017NatPh..13..677F}.

The TLS noise spectrum, Eq.~(\ref{eq:TLSNoise}), vanishes   slowly at large frequencies, $S(\omega) \sim 1/\omega^2$, resulting in a divergent integral~(\ref{eq:deltaGgeneralintermediate})~\cite{toda2012statistical,*forster2018hydrodynamic}. 
Our derivation of  Eq.~(\ref{eq:deltaGgeneralintermediate}) starting from Eq.~(\ref{eq:Hv}) treated the noise source $w$ as classical, which restricts the validity of Eq.~(\ref{eq:deltaGgeneralintermediate}) to low frequencies, $\omega \ll T$. 
If the dominant contribution to the integral comes from higher frequencies, as is the case for the noise spectrum~(\ref{eq:TLSNoise}), one must use an equation that is valid also at higher frequencies. 
Such an equation is obtained from a proper quantum derivation that  treats $w$ as an operator, see Ref.~\cite{Note3}. 
Equation~(\ref{eq:deltaGgeneralintermediate}) generalized to higher frequencies is obtained by replacing $S(\omega) \to \frac{(  \omega / 2T )^2}{\sinh ^2 (  \omega / 2T )} S(\omega)$ in that equation. 
Using this quantum form in Eq.~(\ref{eq:deltaGgeneralintermediate}) yields 
\begin{equation}
\delta G  =G_0 \frac{2 \pi }{3 v^2} |U_{2k_{F}}|^{2}  n_{\perp}^2 \frac{T}{D^{2}\tau} p_{0}(1-p_{0})    \,,\label{eq:singleTLSfinal}
\end{equation} 	
in the limit $\tau^{-1} \ll T$. 
This is the more refined version of Eq.~(\ref{eq:singeTLSqualit}) 
derived in the introduction. 

The temperature-dependence of Eq.~(\ref{eq:singleTLSfinal}) arises from three factors: the puddle occupation number $p_0$, the factor $T$ coming from the sum over frequencies contributing to backscattering, and finally from the so-far unspecified TLS relaxation rate $\tau^{-1}$. 
The relaxation time is a sum of two microscopic times, $\tau = \tau_{esc} + \tau_{e-e}$: the time $\tau_{esc}$ of elastic electron escape  from the puddle and the inelastic   energy relaxation time $\tau_{e-e}$. 
The former is independent of temperature, while the latter for charge puddles varies as~\cite{2014PhRvB..90k5309V} $\tau_{e-e}^{-1} \propto T^2 /\delta$ where $\delta$ is the puddle level spacing. At  temperature much higher than  $\sqrt{\delta \tau_{esc}^{-1}}$ one has $\tau_{e-e}^{-1} \gg  \tau_{esc}^{-1}$ so that $\tau^{-1} \approx \tau_{esc}^{-1}$ is almost independent of temperature. The conductance correction, Eq.~(\ref{eq:singleTLSfinal}), has then rather weak temperature-dependence. We focus on this limit $\tau^{-1} \approx  \tau_{esc}^{-1}$  hereafter.

\textbf{\emph{Long edge.}}~Equation~(\ref{eq:singleTLSfinal}) 
is valid for a single fluctuating TLS which is the 
relevant case for a short edge. Next, we shall consider the effects
of multiple TLSs near the edge, which is appropriate for a long edge.
The correction to conductance, Eq.~(\ref{eq:singleTLSfinal}), due
to a single TLS can be translated into an added small resistance $\delta R=\delta G/G_{0}^{2}\ll G_{0}^{-1}$
to the total edge resistance, $R\approx G_{0}^{-1}+\delta R$. 
Assuming uncorrelated fluctuations of the TLSs, we can neglect interference contributions~\footnote{For example, in the case of two TLSs, of the form Eq.~(\ref{eq:Hvwithtexture}),
the interference term will be proportional to $w_{1}(t)w_{2}(t+t')$,
whose average over $t$ is independent of $t'$ and therefore it doesn't
contribute to $\overline{\langle\delta I\rangle}$ for the same reason
as a static perturbation does not contribute to it. }; the resistance of a long edge is then $R\approx\sum_{i}\delta R_{i}$, 
where $\delta R_{i}$ is the contribution from the $i$th TLS. Summing
over $i$ amounts to ensemble-averaging Eq.~(\ref{eq:singleTLSfinal})
over the random parameters, in particular $N_{g}$. 
The average is dominated by those puddles where $N_{g}$ is close to a half-integer, $\Delta \ll T$.
Interpreting $\tau^{-1}$ and $|U_{2k_{F}}|^{2}$ as their typical
	values at $\{N_g\}\approx 1/2$, we find therefore an edge resistance 
\begin{equation}
R = Ln G_0^{-1} \frac{\pi |U_{2k_{F}}|^{2}}{3 v^2}  n_{\perp}^{2} \frac{T}{D^{2}\tau} \frac{T}{E_{c}}\tanh\frac{E_{c}}{2T} \,,\label{eq:Resistance}
\end{equation}
where $n$ is the one-dimensional impurity density~\footnote{The density $n$ is obtained from the two-dimensional density $n_{2D}$
by multiplying the latter with the effective range $\xi$ of the potential
$U$; $n=n_{2D}\,\xi$. In practice, $\xi$ is the screening length
given by the distance to nearest gate.} and $L$ is the length of the edge. 
The factor $(2T/E_{C})\tanh\frac{E_{c}}{2T} =4 \langle p_{0}(1-p_{0})\rangle_{N_g}$ is the fraction of TLS for which $T\gg \Delta$. 
The noise time scale $\tau$ can be estimated by evaluating the escape rate of an electron from one charge puddle to a neighboring puddle [see comments below Eq.~(\ref{eq:singleTLSfinal})]. 
Since $\{N_g\} \approx 1/2$, tunneling is resonant and its rate is equal to the
level splitting in the two-puddle problem. 
We can use the WKB approximation since the puddles are typically large~\cite{2014PhRvB..90k5309V}. 
This estimate gives $\tau^{-1}\sim\delta e^{-d/\Lambda}$, where $\delta$
is the puddle level spacing, $\Lambda$ is the penetration depth of
a low-energy electron into the bulk and $d$ is the average distance
between the puddles. 
In the limit $T \gg E_{c}$ the resistance, Eq.~(\ref{eq:Resistance}), is linear in temperature, while at low temperatures $R\propto T^2$. 
Strictly speaking, at $T \gg E_{c}$ one should include more charge states in our noise model. 
As long as $\tau^{-1} \ll E_c$, the puddle charge takes discrete well-defined values and the noise has the Lorentzian form, Eq.~(\ref{eq:TLSNoise}). 
The generalization of the prefactor $p_0 (1-p_0)$ in that equation to many levels remains $T$-independent at high temperatures.

\textbf{\emph{$1/f$ noise.}}~Electronics ubiquitously exhibit $1/f$ noise.  We therefore discuss the resulting resistance for this case, assuming that the noise source is extensive. 
Let us first discuss in which frequency range the noise from a collection of charge puddles can have a $1/f$ form. 
In Eq.~(\ref{eq:Resistance}) we took $\tau^{-1}\sim\delta e^{-d/\Lambda}$
with $d$ the puddle-puddle distance. Assuming that the random variable
$d$ is distributed uniformly, the corresponding distribution of relaxation
times, $P(\tau)\sim1/\tau$, is dominated by short times $\tau\sim1/\delta$.
Further assuming $\delta \ll T$, the resistance averaged over puddle positions then becomes  
\begin{equation}
R \sim Ln G_0^{-1} \frac{T \delta |U_{2k_{F}}|^{2}}{ D^2 v^2}  n_{\perp}^{2} \,,\label{eq:resistance2}
\end{equation}
in the high-temperature limit $T \gg E_c$ while $R \propto T^2$ at $T \ll E_c$. 
By averaging Eq.~(\ref{eq:TLSNoise}) over $\tau$ with the distribution $1/\tau$, the resulting noise spectrum is $1/\omega$ at low frequencies, $\omega \ll  \delta$, but $1/\omega^2$ at high frequencies $\omega \gg  \delta$. When $\delta \ll T$, the high-frequency part gives the dominant contribution to Eq.~(\ref{eq:resistance2}). 
This is because the transition matrix element    becomes small at low energy transfers, $[B_{k'+k_{F}}^{\dagger}B_{k+k_{F}}]_{10} \approx \delta /D $ when $v(k'-k) \approx \delta \ll D$; see discussion above Eq.~(\ref{eq:deltaGgeneralintermediate}).

Let us next discuss the more generic $1/f^\gamma$ noise without specifying its microscopic origin. 
For $1/f^{\gamma}$ noise ($0<\gamma <3$) with a \textit{sharp} high-frequency cutoff $\Omega \ll T$ we find  $R \sim   G_0^{-1} (\Omega / D)^{2}  n_{\perp}^{2} S_0  |U_{2k_{F}}|^{2} /v^2$ with an $\gamma$-dependent numerical coefficient. 
This result is obtained from  Eq.~(\ref{eq:deltaGgeneral}) by taking  a spectrum $S(\omega)=|\omega|^{-\gamma}\Omega^{\gamma-1}S_{0}$, defined for $|\omega|\!<\Omega$.

%
%

\textbf{\emph{Discussion.}}~The noise-induced backscattering underlying  Eq.~(\ref{eq:deltaGgeneral}) relies on the presence
of a momentum-space spin texture of the edge state. Although band
theory predicts the existence of such a texture~\cite{PhysRevB.91.245112},
its experimental detection is so far absent. 
As a simple experimental probe of the spin texture, we suggest creating a time-dependent ``noise potential'' artificially by an external gate; see Fig.~\ref{fig:1}b. With ac voltage $V_{0}\cos\omega_{0}t$
applied to the gate, we have $U(x)=V_{0}u(x)$ and $w(t)=\cos\omega_{0}t$ in Eq.~(\ref{eq:Hv}). 
The dimensionless function $u(x)$ is geometry-dependent
and can be in principle found by solving the Poisson equation. 
Using Eq.~(\ref{eq:deltaGgeneralintermediate}) with $S(\omega)$ 
obtained from
$w$,  we find in the limit $\omega_{0}\,, T\ll D,\,\mu$, 
\begin{equation}
\delta G  = G_0 \frac{ \omega_{0}^{2}}{4 D^{2} v^2} V_0^2 |u_{2k_{F}}|^{2}  n_{\perp}^2  \,.\label{eq:deltaGexp}
\end{equation}
Remarkably, by the application of an ac gate voltage, one may create
an effective backscattering potential $\propto\omega_{0}V_{0}u_{2k_{F}} n_\perp /D$
on the helical edge, as long as a spin texture exists. Quadratic dependence of $\delta G$ on gate voltage and frequency  thus constitutes a clear experimental signature of a spin texture in a helical edge. 
We note that for large $k_{F}=\mu/v$ it may be difficult in practice to create a sharp
enough potential that induces substantial $u_{2k_{F}}$. This difficulty can
be avoided if the Dirac point is not buried~\cite{Wu76,2017arXiv170904830S}
and one can tune $\mu $ with a global gate to a smaller value. In
this limit there is an additional $\mu$-dependence in $\delta G$
stemming from $D^{-2} \propto \mu^2$; see discussion below Eq.~(\ref{eq:deltaGgeneralintermediate}). 

Let us finally estimate parameters of our noise models and discuss the size of the effect. 
For charge puddles, as was mentioned in the context of Eq.~(\ref{eq:resistance2}), 
the dominant contribution to resistance comes from close pairs of
puddles for which $\tau^{-1}\approx\delta$. For HgTe, $ \delta \approx \alpha^2 \Delta_{b}$ with $\alpha=e^{2}/ 4\pi \epsilon v $ and $\Delta_{b}$ is the band gap~\footnote{From Ref.~\cite{2014PhRvB..90k5309V} we have $\delta\approx\alpha^2 \Delta_{b}$ 	and $\alpha\approx0.3$. }. 
For a ball-park estimate of $U_{2k_{F}}$, we can use the charging
energy $E_{c}\sim\alpha v/l$ of a charge puddle of size $l$. Assuming a short-range potential then gives
$U_{2k_{F}}\sim v\alpha$. Taking $n_\perp \sim1$, we obtain the final estimate for the resistance
in Eq.~(\ref{eq:resistance2}), $R \sim Ln G_0^{-1} \frac{\alpha^{4} T \Delta_{b}}{ D^2} $. 
In HgTe, the interaction constant $\alpha \approx 0.3$ is not very large and near-edge puddles are possibly rare, $n \sim 1/\mu m$. 
Therefore, in HgTe fluctuating dipoles (modeled as TLSs) in the dielectric may give a larger source of resistance~\cite{Note3}. 
This contribution is  $R \sim \frac{1}{G_{0}} \frac{LT}{v} \alpha N e^{-4d_0 k_{F}} \frac{T^{2}}{D^{2}}\tan\delta$, where $N$ is the number of TLS in the dielectric and   $\tan\delta$ is its loss tangent; $d_0$ is the distance of the dielectric from the edge~\footnote{For TLS in the bulk of the 2D TI, one can set $d_0 \to 0$ for TLS that are very close to the edge.}. 
For example, a  $\text{SiO}_2$ dielectric of size $2\times 2 \times 0.1\, \mu m^3$ ($L=2\mu m$)  at temperature $T=1$K  hosts~\cite{PhysRevB.8.2896} $N \sim 2 \times 10^{4}$ contributing TLSs and has a loss tangent~\cite{PhysRevB.79.094520}   $\tan\delta \sim 10^{-3}$. Assuming $T/D \sim 1$ and focusing on  the vicinity of the Dirac point, $k_F d_0 \lesssim 1$, the corresponding resistance is significant, $R \sim  G_0^{-1}$, even for a short $L=2\mu m$ edge. 
For even shorter edges the dominant contribution may come from gate noise, see Eq.~(\ref{eq:deltaGexp}).

Our main results, Eqs.~(\ref{eq:singleTLSfinal})\textendash (\ref{eq:Resistance}),
and the above estimates were derived for a specific model of a fluctuating
dipole in thermal equilibrium. We emphasize that these results generalize
to the case of a non-equilibrium noise source. An important example
is when the effective noise temperature is much higher than the system
temperature. The result for that case can be obtained by taking the
high temperature limit 
in Eqs.~(\ref{eq:singleTLSfinal})\textendash (\ref{eq:Resistance}). 

The mechanism of noise-induced backscattering may play a role in broader settings in materials where elastic backscattering is suppressed, for example in graphene or in 3D topological insulators.  
Future studies of noise in such context may extend to topics such as spin relaxation~\cite{garcia2018spin} and dephasing of quasiparticle  interference~\cite{PhysRevLett.103.266803,*PhysRevLett.104.016401}.  



%
{\bf \emph{Acknowledgments.}}~We thank  Gijs de Lange and John Watson for helpful discussions and especially thank Leonid Glazman for his perceptive insights. We also gratefully acknowledge support from the National
Science Foundation through grant DMR-1723367; the Caltech Institute
for Quantum Information and Matter, an NSF Physics
Frontiers Center with support of the Gordon and Betty
Moore Foundation through Grant GBMF1250; and the
Walter Burke Institute for Theoretical Physics at Caltech.

%

\bibliographystyle{apsrev4-1}
\bibliography{refs}

\begin{thebibliography}{48}%
\makeatletter
\providecommand \@ifxundefined [1]{%
 \@ifx{#1\undefined}
}%
\providecommand \@ifnum [1]{%
 \ifnum #1\expandafter \@firstoftwo
 \else \expandafter \@secondoftwo
 \fi
}%
\providecommand \@ifx [1]{%
 \ifx #1\expandafter \@firstoftwo
 \else \expandafter \@secondoftwo
 \fi
}%
\providecommand \natexlab [1]{#1}%
\providecommand \enquote  [1]{``#1''}%
\providecommand \bibnamefont  [1]{#1}%
\providecommand \bibfnamefont [1]{#1}%
\providecommand \citenamefont [1]{#1}%
\providecommand \href@noop [0]{\@secondoftwo}%
\providecommand \href [0]{\begingroup \@sanitize@url \@href}%
\providecommand \@href[1]{\@@startlink{#1}\@@href}%
\providecommand \@@href[1]{\endgroup#1\@@endlink}%
\providecommand \@sanitize@url [0]{\catcode `\\12\catcode `\$12\catcode
  `\&12\catcode `\#12\catcode `\^12\catcode `\_12\catcode `\%12\relax}%
\providecommand \@@startlink[1]{}%
\providecommand \@@endlink[0]{}%
\providecommand \url  [0]{\begingroup\@sanitize@url \@url }%
\providecommand \@url [1]{\endgroup\@href {#1}{\urlprefix }}%
\providecommand \urlprefix  [0]{URL }%
\providecommand \Eprint [0]{\href }%
\providecommand \doibase [0]{http://dx.doi.org/}%
\providecommand \selectlanguage [0]{\@gobble}%
\providecommand \bibinfo  [0]{\@secondoftwo}%
\providecommand \bibfield  [0]{\@secondoftwo}%
\providecommand \translation [1]{[#1]}%
\providecommand \BibitemOpen [0]{}%
\providecommand \bibitemStop [0]{}%
\providecommand \bibitemNoStop [0]{.\EOS\space}%
\providecommand \EOS [0]{\spacefactor3000\relax}%
\providecommand \BibitemShut  [1]{\csname bibitem#1\endcsname}%
\let\auto@bib@innerbib\@empty
\bibitem [{\citenamefont {{Kane}}\ and\ \citenamefont
  {{Mele}}(2005{\natexlab{a}})}]{2005PhRvL..95n6802K}%
  \BibitemOpen
  \bibfield  {author} {\bibinfo {author} {\bibfnamefont {C.~L.}\ \bibnamefont
  {{Kane}}}\ and\ \bibinfo {author} {\bibfnamefont {E.~J.}\ \bibnamefont
  {{Mele}}},\ }\href {\doibase 10.1103/PhysRevLett.95.146802} {\bibfield
  {journal} {\bibinfo  {journal} {Physical Review Letters}\ }\textbf {\bibinfo
  {volume} {95}},\ \bibinfo {eid} {146802} (\bibinfo {year}
  {2005}{\natexlab{a}})},\ \Eprint {http://arxiv.org/abs/cond-mat/0506581}
  {cond-mat/0506581} \BibitemShut {NoStop}%
\bibitem [{\citenamefont {{Kane}}\ and\ \citenamefont
  {{Mele}}(2005{\natexlab{b}})}]{2005PhRvL..95v6801K}%
  \BibitemOpen
  \bibfield  {author} {\bibinfo {author} {\bibfnamefont {C.~L.}\ \bibnamefont
  {{Kane}}}\ and\ \bibinfo {author} {\bibfnamefont {E.~J.}\ \bibnamefont
  {{Mele}}},\ }\href {\doibase 10.1103/PhysRevLett.95.226801} {\bibfield
  {journal} {\bibinfo  {journal} {Physical Review Letters}\ }\textbf {\bibinfo
  {volume} {95}},\ \bibinfo {eid} {226801} (\bibinfo {year}
  {2005}{\natexlab{b}})},\ \Eprint {http://arxiv.org/abs/cond-mat/0411737}
  {cond-mat/0411737} \BibitemShut {NoStop}%
\bibitem [{\citenamefont {{Bernevig}}\ and\ \citenamefont
  {{Zhang}}(2006)}]{2006PhRvL..96j6802B}%
  \BibitemOpen
  \bibfield  {author} {\bibinfo {author} {\bibfnamefont {B.~A.}\ \bibnamefont
  {{Bernevig}}}\ and\ \bibinfo {author} {\bibfnamefont {S.-C.}\ \bibnamefont
  {{Zhang}}},\ }\href {\doibase 10.1103/PhysRevLett.96.106802} {\bibfield
  {journal} {\bibinfo  {journal} {Physical Review Letters}\ }\textbf {\bibinfo
  {volume} {96}},\ \bibinfo {eid} {106802} (\bibinfo {year} {2006})},\ \Eprint
  {http://arxiv.org/abs/cond-mat/0504147} {cond-mat/0504147} \BibitemShut
  {NoStop}%
\bibitem [{\citenamefont {{K{\"o}nig}}\ \emph {et~al.}(2007)\citenamefont
  {{K{\"o}nig}}, \citenamefont {{Wiedmann}}, \citenamefont {{Br{\"u}ne}},
  \citenamefont {{Roth}}, \citenamefont {{Buhmann}}, \citenamefont
  {{Molenkamp}}, \citenamefont {{Qi}},\ and\ \citenamefont
  {{Zhang}}}]{2007Sci...318..766K}%
  \BibitemOpen
  \bibfield  {author} {\bibinfo {author} {\bibfnamefont {M.}~\bibnamefont
  {{K{\"o}nig}}}, \bibinfo {author} {\bibfnamefont {S.}~\bibnamefont
  {{Wiedmann}}}, \bibinfo {author} {\bibfnamefont {C.}~\bibnamefont
  {{Br{\"u}ne}}}, \bibinfo {author} {\bibfnamefont {A.}~\bibnamefont {{Roth}}},
  \bibinfo {author} {\bibfnamefont {H.}~\bibnamefont {{Buhmann}}}, \bibinfo
  {author} {\bibfnamefont {L.~W.}\ \bibnamefont {{Molenkamp}}}, \bibinfo
  {author} {\bibfnamefont {X.-L.}\ \bibnamefont {{Qi}}}, \ and\ \bibinfo
  {author} {\bibfnamefont {S.-C.}\ \bibnamefont {{Zhang}}},\ }\href {\doibase
  10.1126/science.1148047} {\bibfield  {journal} {\bibinfo  {journal}
  {Science}\ }\textbf {\bibinfo {volume} {318}},\ \bibinfo {pages} {766}
  (\bibinfo {year} {2007})},\ \Eprint {http://arxiv.org/abs/0710.0582}
  {arXiv:0710.0582} \BibitemShut {NoStop}%
\bibitem [{\citenamefont {{Roth}}\ \emph {et~al.}(2009)\citenamefont {{Roth}},
  \citenamefont {{Br{\"u}ne}}, \citenamefont {{Buhmann}}, \citenamefont
  {{Molenkamp}}, \citenamefont {{Maciejko}}, \citenamefont {{Qi}},\ and\
  \citenamefont {{Zhang}}}]{2009Sci...325..294R}%
  \BibitemOpen
  \bibfield  {author} {\bibinfo {author} {\bibfnamefont {A.}~\bibnamefont
  {{Roth}}}, \bibinfo {author} {\bibfnamefont {C.}~\bibnamefont {{Br{\"u}ne}}},
  \bibinfo {author} {\bibfnamefont {H.}~\bibnamefont {{Buhmann}}}, \bibinfo
  {author} {\bibfnamefont {L.~W.}\ \bibnamefont {{Molenkamp}}}, \bibinfo
  {author} {\bibfnamefont {J.}~\bibnamefont {{Maciejko}}}, \bibinfo {author}
  {\bibfnamefont {X.-L.}\ \bibnamefont {{Qi}}}, \ and\ \bibinfo {author}
  {\bibfnamefont {S.-C.}\ \bibnamefont {{Zhang}}},\ }\href {\doibase
  10.1126/science.1174736} {\bibfield  {journal} {\bibinfo  {journal}
  {Science}\ }\textbf {\bibinfo {volume} {325}},\ \bibinfo {pages} {294}
  (\bibinfo {year} {2009})},\ \Eprint {http://arxiv.org/abs/0905.0365}
  {arXiv:0905.0365} \BibitemShut {NoStop}%
\bibitem [{\citenamefont {{Gusev}}\ \emph {et~al.}(2014)\citenamefont
  {{Gusev}}, \citenamefont {{Kvon}}, \citenamefont {{Olshanetsky}},
  \citenamefont {{Levin}}, \citenamefont {{Krupko}}, \citenamefont {{Portal}},
  \citenamefont {{Mikhailov}},\ and\ \citenamefont
  {{Dvoretsky}}}]{2014PhRvB..89l5305G}%
  \BibitemOpen
  \bibfield  {author} {\bibinfo {author} {\bibfnamefont {G.~M.}\ \bibnamefont
  {{Gusev}}}, \bibinfo {author} {\bibfnamefont {Z.~D.}\ \bibnamefont {{Kvon}}},
  \bibinfo {author} {\bibfnamefont {E.~B.}\ \bibnamefont {{Olshanetsky}}},
  \bibinfo {author} {\bibfnamefont {A.~D.}\ \bibnamefont {{Levin}}}, \bibinfo
  {author} {\bibfnamefont {Y.}~\bibnamefont {{Krupko}}}, \bibinfo {author}
  {\bibfnamefont {J.~C.}\ \bibnamefont {{Portal}}}, \bibinfo {author}
  {\bibfnamefont {N.~N.}\ \bibnamefont {{Mikhailov}}}, \ and\ \bibinfo {author}
  {\bibfnamefont {S.~A.}\ \bibnamefont {{Dvoretsky}}},\ }\href {\doibase
  10.1103/PhysRevB.89.125305} {\bibfield  {journal} {\bibinfo  {journal}
  {\prb}\ }\textbf {\bibinfo {volume} {89}},\ \bibinfo {eid} {125305} (\bibinfo
  {year} {2014})}\BibitemShut {NoStop}%
\bibitem [{\citenamefont {Knez}\ \emph {et~al.}(2014)\citenamefont {Knez},
  \citenamefont {Rettner}, \citenamefont {Yang}, \citenamefont {Parkin},
  \citenamefont {Du}, \citenamefont {Du},\ and\ \citenamefont
  {Sullivan}}]{Knez14}%
  \BibitemOpen
  \bibfield  {author} {\bibinfo {author} {\bibfnamefont {I.}~\bibnamefont
  {Knez}}, \bibinfo {author} {\bibfnamefont {C.~T.}\ \bibnamefont {Rettner}},
  \bibinfo {author} {\bibfnamefont {S.-H.}\ \bibnamefont {Yang}}, \bibinfo
  {author} {\bibfnamefont {S.~S.~P.}\ \bibnamefont {Parkin}}, \bibinfo {author}
  {\bibfnamefont {L.}~\bibnamefont {Du}}, \bibinfo {author} {\bibfnamefont
  {R.-R.}\ \bibnamefont {Du}}, \ and\ \bibinfo {author} {\bibfnamefont
  {G.}~\bibnamefont {Sullivan}},\ }\href {\doibase
  10.1103/PhysRevLett.112.026602} {\bibfield  {journal} {\bibinfo  {journal}
  {Phys. Rev. Lett.}\ }\textbf {\bibinfo {volume} {112}},\ \bibinfo {pages}
  {026602} (\bibinfo {year} {2014})}\BibitemShut {NoStop}%
\bibitem [{\citenamefont {Du}\ \emph {et~al.}(2015)\citenamefont {Du},
  \citenamefont {Knez}, \citenamefont {Sullivan},\ and\ \citenamefont
  {Du}}]{Du2015}%
  \BibitemOpen
  \bibfield  {author} {\bibinfo {author} {\bibfnamefont {L.}~\bibnamefont
  {Du}}, \bibinfo {author} {\bibfnamefont {I.}~\bibnamefont {Knez}}, \bibinfo
  {author} {\bibfnamefont {G.}~\bibnamefont {Sullivan}}, \ and\ \bibinfo
  {author} {\bibfnamefont {R.-R.}\ \bibnamefont {Du}},\ }\href {\doibase
  10.1103/PhysRevLett.114.096802} {\bibfield  {journal} {\bibinfo  {journal}
  {Phys. Rev. Lett.}\ }\textbf {\bibinfo {volume} {114}},\ \bibinfo {pages}
  {096802} (\bibinfo {year} {2015})}\BibitemShut {NoStop}%
\bibitem [{\citenamefont {Nichele}\ \emph {et~al.}(2016)\citenamefont
  {Nichele}, \citenamefont {Suominen}, \citenamefont {Kjaergaard},
  \citenamefont {Marcus}, \citenamefont {Sajadi}, \citenamefont {Folk},
  \citenamefont {Qu}, \citenamefont {Beukman}, \citenamefont {de~Vries},
  \citenamefont {van Veen}, \citenamefont {Nadj-Perge}, \citenamefont
  {Kouwenhoven}, \citenamefont {Nguyen}, \citenamefont {Kiselev}, \citenamefont
  {Yi}, \citenamefont {Sokolich}, \citenamefont {Manfra}, \citenamefont
  {Spanton},\ and\ \citenamefont {Moler}}]{1367-2630-18-8-083005}%
  \BibitemOpen
  \bibfield  {author} {\bibinfo {author} {\bibfnamefont {F.}~\bibnamefont
  {Nichele}}, \bibinfo {author} {\bibfnamefont {H.~J.}\ \bibnamefont
  {Suominen}}, \bibinfo {author} {\bibfnamefont {M.}~\bibnamefont
  {Kjaergaard}}, \bibinfo {author} {\bibfnamefont {C.~M.}\ \bibnamefont
  {Marcus}}, \bibinfo {author} {\bibfnamefont {E.}~\bibnamefont {Sajadi}},
  \bibinfo {author} {\bibfnamefont {J.~A.}\ \bibnamefont {Folk}}, \bibinfo
  {author} {\bibfnamefont {F.}~\bibnamefont {Qu}}, \bibinfo {author}
  {\bibfnamefont {A.~J.~A.}\ \bibnamefont {Beukman}}, \bibinfo {author}
  {\bibfnamefont {F.~K.}\ \bibnamefont {de~Vries}}, \bibinfo {author}
  {\bibfnamefont {J.}~\bibnamefont {van Veen}}, \bibinfo {author}
  {\bibfnamefont {S.}~\bibnamefont {Nadj-Perge}}, \bibinfo {author}
  {\bibfnamefont {L.~P.}\ \bibnamefont {Kouwenhoven}}, \bibinfo {author}
  {\bibfnamefont {B.-M.}\ \bibnamefont {Nguyen}}, \bibinfo {author}
  {\bibfnamefont {A.~A.}\ \bibnamefont {Kiselev}}, \bibinfo {author}
  {\bibfnamefont {W.}~\bibnamefont {Yi}}, \bibinfo {author} {\bibfnamefont
  {M.}~\bibnamefont {Sokolich}}, \bibinfo {author} {\bibfnamefont {M.~J.}\
  \bibnamefont {Manfra}}, \bibinfo {author} {\bibfnamefont {E.~M.}\
  \bibnamefont {Spanton}}, \ and\ \bibinfo {author} {\bibfnamefont {K.~A.}\
  \bibnamefont {Moler}},\ }\href
  {http://stacks.iop.org/1367-2630/18/i=8/a=083005} {\bibfield  {journal}
  {\bibinfo  {journal} {New Journal of Physics}\ }\textbf {\bibinfo {volume}
  {18}},\ \bibinfo {pages} {083005} (\bibinfo {year} {2016})}\BibitemShut
  {NoStop}%
\bibitem [{\citenamefont {Du}\ \emph {et~al.}(2017)\citenamefont {Du},
  \citenamefont {Li}, \citenamefont {Lou}, \citenamefont {Wu}, \citenamefont
  {Liu}, \citenamefont {Han}, \citenamefont {Zhang}, \citenamefont {Sullivan},
  \citenamefont {Ikhlassi}, \citenamefont {Chang},\ and\ \citenamefont
  {Du}}]{PhysRevLett.119.056803}%
  \BibitemOpen
  \bibfield  {author} {\bibinfo {author} {\bibfnamefont {L.}~\bibnamefont
  {Du}}, \bibinfo {author} {\bibfnamefont {T.}~\bibnamefont {Li}}, \bibinfo
  {author} {\bibfnamefont {W.}~\bibnamefont {Lou}}, \bibinfo {author}
  {\bibfnamefont {X.}~\bibnamefont {Wu}}, \bibinfo {author} {\bibfnamefont
  {X.}~\bibnamefont {Liu}}, \bibinfo {author} {\bibfnamefont {Z.}~\bibnamefont
  {Han}}, \bibinfo {author} {\bibfnamefont {C.}~\bibnamefont {Zhang}}, \bibinfo
  {author} {\bibfnamefont {G.}~\bibnamefont {Sullivan}}, \bibinfo {author}
  {\bibfnamefont {A.}~\bibnamefont {Ikhlassi}}, \bibinfo {author}
  {\bibfnamefont {K.}~\bibnamefont {Chang}}, \ and\ \bibinfo {author}
  {\bibfnamefont {R.-R.}\ \bibnamefont {Du}},\ }\href {\doibase
  10.1103/PhysRevLett.119.056803} {\bibfield  {journal} {\bibinfo  {journal}
  {Phys. Rev. Lett.}\ }\textbf {\bibinfo {volume} {119}},\ \bibinfo {pages}
  {056803} (\bibinfo {year} {2017})}\BibitemShut {NoStop}%
\bibitem [{\citenamefont {{Fei}}\ \emph {et~al.}(2017)\citenamefont {{Fei}},
  \citenamefont {{Palomaki}}, \citenamefont {{Wu}}, \citenamefont {{Zhao}},
  \citenamefont {{Cai}}, \citenamefont {{Sun}}, \citenamefont {{Nguyen}},
  \citenamefont {{Finney}}, \citenamefont {{Xu}},\ and\ \citenamefont
  {{Cobden}}}]{2017NatPh..13..677F}%
  \BibitemOpen
  \bibfield  {author} {\bibinfo {author} {\bibfnamefont {Z.}~\bibnamefont
  {{Fei}}}, \bibinfo {author} {\bibfnamefont {T.}~\bibnamefont {{Palomaki}}},
  \bibinfo {author} {\bibfnamefont {S.}~\bibnamefont {{Wu}}}, \bibinfo {author}
  {\bibfnamefont {W.}~\bibnamefont {{Zhao}}}, \bibinfo {author} {\bibfnamefont
  {X.}~\bibnamefont {{Cai}}}, \bibinfo {author} {\bibfnamefont
  {B.}~\bibnamefont {{Sun}}}, \bibinfo {author} {\bibfnamefont
  {P.}~\bibnamefont {{Nguyen}}}, \bibinfo {author} {\bibfnamefont
  {J.}~\bibnamefont {{Finney}}}, \bibinfo {author} {\bibfnamefont
  {X.}~\bibnamefont {{Xu}}}, \ and\ \bibinfo {author} {\bibfnamefont {D.~H.}\
  \bibnamefont {{Cobden}}},\ }\href {\doibase 10.1038/nphys4091} {\bibfield
  {journal} {\bibinfo  {journal} {Nature Physics}\ }\textbf {\bibinfo {volume}
  {13}},\ \bibinfo {pages} {677} (\bibinfo {year} {2017})},\ \Eprint
  {http://arxiv.org/abs/1610.07924} {arXiv:1610.07924} \BibitemShut {NoStop}%
\bibitem [{\citenamefont {Wu}\ \emph {et~al.}(2018)\citenamefont {Wu},
  \citenamefont {Fatemi}, \citenamefont {Gibson}, \citenamefont {Watanabe},
  \citenamefont {Taniguchi}, \citenamefont {Cava},\ and\ \citenamefont
  {Jarillo-Herrero}}]{Wu76}%
  \BibitemOpen
  \bibfield  {author} {\bibinfo {author} {\bibfnamefont {S.}~\bibnamefont
  {Wu}}, \bibinfo {author} {\bibfnamefont {V.}~\bibnamefont {Fatemi}}, \bibinfo
  {author} {\bibfnamefont {Q.~D.}\ \bibnamefont {Gibson}}, \bibinfo {author}
  {\bibfnamefont {K.}~\bibnamefont {Watanabe}}, \bibinfo {author}
  {\bibfnamefont {T.}~\bibnamefont {Taniguchi}}, \bibinfo {author}
  {\bibfnamefont {R.~J.}\ \bibnamefont {Cava}}, \ and\ \bibinfo {author}
  {\bibfnamefont {P.}~\bibnamefont {Jarillo-Herrero}},\ }\href {\doibase
  10.1126/science.aan6003} {\bibfield  {journal} {\bibinfo  {journal}
  {Science}\ }\textbf {\bibinfo {volume} {359}},\ \bibinfo {pages} {76}
  (\bibinfo {year} {2018})},\ \Eprint
  {http://arxiv.org/abs/http://science.sciencemag.org/content/359/6371/76.full.pdf}
  {http://science.sciencemag.org/content/359/6371/76.full.pdf} \BibitemShut
  {NoStop}%
\bibitem [{\citenamefont {{Xu}}\ and\ \citenamefont
  {{Moore}}(2006)}]{2006PhRvB..73d5322X}%
  \BibitemOpen
  \bibfield  {author} {\bibinfo {author} {\bibfnamefont {C.}~\bibnamefont
  {{Xu}}}\ and\ \bibinfo {author} {\bibfnamefont {J.~E.}\ \bibnamefont
  {{Moore}}},\ }\href {\doibase 10.1103/PhysRevB.73.045322} {\bibfield
  {journal} {\bibinfo  {journal} {\prb}\ }\textbf {\bibinfo {volume} {73}},\
  \bibinfo {eid} {045322} (\bibinfo {year} {2006})},\ \Eprint
  {http://arxiv.org/abs/cond-mat/0508291} {cond-mat/0508291} \BibitemShut
  {NoStop}%
\bibitem [{\citenamefont {{Wu}}\ \emph {et~al.}(2006)\citenamefont {{Wu}},
  \citenamefont {{Bernevig}},\ and\ \citenamefont
  {{Zhang}}}]{2006PhRvL..96j6401W}%
  \BibitemOpen
  \bibfield  {author} {\bibinfo {author} {\bibfnamefont {C.}~\bibnamefont
  {{Wu}}}, \bibinfo {author} {\bibfnamefont {B.~A.}\ \bibnamefont
  {{Bernevig}}}, \ and\ \bibinfo {author} {\bibfnamefont {S.-C.}\ \bibnamefont
  {{Zhang}}},\ }\href {\doibase 10.1103/PhysRevLett.96.106401} {\bibfield
  {journal} {\bibinfo  {journal} {Physical Review Letters}\ }\textbf {\bibinfo
  {volume} {96}},\ \bibinfo {eid} {106401} (\bibinfo {year} {2006})},\ \Eprint
  {http://arxiv.org/abs/cond-mat/0508273} {cond-mat/0508273} \BibitemShut
  {NoStop}%
\bibitem [{\citenamefont {{Maciejko}}\ \emph {et~al.}(2009)\citenamefont
  {{Maciejko}}, \citenamefont {{Liu}}, \citenamefont {{Oreg}}, \citenamefont
  {{Qi}}, \citenamefont {{Wu}},\ and\ \citenamefont
  {{Zhang}}}]{2009PhRvL.102y6803M}%
  \BibitemOpen
  \bibfield  {author} {\bibinfo {author} {\bibfnamefont {J.}~\bibnamefont
  {{Maciejko}}}, \bibinfo {author} {\bibfnamefont {C.}~\bibnamefont {{Liu}}},
  \bibinfo {author} {\bibfnamefont {Y.}~\bibnamefont {{Oreg}}}, \bibinfo
  {author} {\bibfnamefont {X.-L.}\ \bibnamefont {{Qi}}}, \bibinfo {author}
  {\bibfnamefont {C.}~\bibnamefont {{Wu}}}, \ and\ \bibinfo {author}
  {\bibfnamefont {S.-C.}\ \bibnamefont {{Zhang}}},\ }\href {\doibase
  10.1103/PhysRevLett.102.256803} {\bibfield  {journal} {\bibinfo  {journal}
  {Physical Review Letters}\ }\textbf {\bibinfo {volume} {102}},\ \bibinfo
  {eid} {256803} (\bibinfo {year} {2009})},\ \Eprint
  {http://arxiv.org/abs/0901.1685} {arXiv:0901.1685} \BibitemShut {NoStop}%
\bibitem [{\citenamefont {{Tanaka}}\ \emph {et~al.}(2011)\citenamefont
  {{Tanaka}}, \citenamefont {{Furusaki}},\ and\ \citenamefont
  {{Matveev}}}]{2011PhRvL.106w6402T}%
  \BibitemOpen
  \bibfield  {author} {\bibinfo {author} {\bibfnamefont {Y.}~\bibnamefont
  {{Tanaka}}}, \bibinfo {author} {\bibfnamefont {A.}~\bibnamefont
  {{Furusaki}}}, \ and\ \bibinfo {author} {\bibfnamefont {K.~A.}\ \bibnamefont
  {{Matveev}}},\ }\href {\doibase 10.1103/PhysRevLett.106.236402} {\bibfield
  {journal} {\bibinfo  {journal} {Physical Review Letters}\ }\textbf {\bibinfo
  {volume} {106}},\ \bibinfo {eid} {236402} (\bibinfo {year} {2011})},\ \Eprint
  {http://arxiv.org/abs/1102.4629} {arXiv:1102.4629} \BibitemShut {NoStop}%
\bibitem [{\citenamefont {{Budich}}\ \emph {et~al.}(2012)\citenamefont
  {{Budich}}, \citenamefont {{Dolcini}}, \citenamefont {{Recher}},\ and\
  \citenamefont {{Trauzettel}}}]{2012PhRvL.108h6602B}%
  \BibitemOpen
  \bibfield  {author} {\bibinfo {author} {\bibfnamefont {J.~C.}\ \bibnamefont
  {{Budich}}}, \bibinfo {author} {\bibfnamefont {F.}~\bibnamefont {{Dolcini}}},
  \bibinfo {author} {\bibfnamefont {P.}~\bibnamefont {{Recher}}}, \ and\
  \bibinfo {author} {\bibfnamefont {B.}~\bibnamefont {{Trauzettel}}},\ }\href
  {\doibase 10.1103/PhysRevLett.108.086602} {\bibfield  {journal} {\bibinfo
  {journal} {Physical Review Letters}\ }\textbf {\bibinfo {volume} {108}},\
  \bibinfo {eid} {086602} (\bibinfo {year} {2012})},\ \Eprint
  {http://arxiv.org/abs/1109.5188} {arXiv:1109.5188} \BibitemShut {NoStop}%
\bibitem [{\citenamefont {Schmidt}\ \emph {et~al.}(2012)\citenamefont
  {Schmidt}, \citenamefont {Rachel}, \citenamefont {von Oppen},\ and\
  \citenamefont {Glazman}}]{schmidt_inelastic_2012}%
  \BibitemOpen
  \bibfield  {author} {\bibinfo {author} {\bibfnamefont {T.~L.}\ \bibnamefont
  {Schmidt}}, \bibinfo {author} {\bibfnamefont {S.}~\bibnamefont {Rachel}},
  \bibinfo {author} {\bibfnamefont {F.}~\bibnamefont {von Oppen}}, \ and\
  \bibinfo {author} {\bibfnamefont {L.~I.}\ \bibnamefont {Glazman}},\ }\href
  {\doibase 10.1103/PhysRevLett.108.156402} {\bibfield  {journal} {\bibinfo
  {journal} {Phys. Rev. Lett.}\ }\textbf {\bibinfo {volume} {108}},\ \bibinfo
  {pages} {156402} (\bibinfo {year} {2012})}\BibitemShut {NoStop}%
\bibitem [{\citenamefont {{Cheianov}}\ and\ \citenamefont
  {{Glazman}}(2013)}]{2013PhRvL.110t6803C}%
  \BibitemOpen
  \bibfield  {author} {\bibinfo {author} {\bibfnamefont {V.}~\bibnamefont
  {{Cheianov}}}\ and\ \bibinfo {author} {\bibfnamefont {L.~I.}\ \bibnamefont
  {{Glazman}}},\ }\href {\doibase 10.1103/PhysRevLett.110.206803} {\bibfield
  {journal} {\bibinfo  {journal} {Physical Review Letters}\ }\textbf {\bibinfo
  {volume} {110}},\ \bibinfo {eid} {206803} (\bibinfo {year} {2013})},\ \Eprint
  {http://arxiv.org/abs/1301.1185} {arXiv:1301.1185} \BibitemShut {NoStop}%
\bibitem [{\citenamefont {{V{\"a}yrynen}}\ \emph {et~al.}(2013)\citenamefont
  {{V{\"a}yrynen}}, \citenamefont {{Goldstein}},\ and\ \citenamefont
  {{Glazman}}}]{2013PhRvL.110u6402V}%
  \BibitemOpen
  \bibfield  {author} {\bibinfo {author} {\bibfnamefont {J.~I.}\ \bibnamefont
  {{V{\"a}yrynen}}}, \bibinfo {author} {\bibfnamefont {M.}~\bibnamefont
  {{Goldstein}}}, \ and\ \bibinfo {author} {\bibfnamefont {L.~I.}\ \bibnamefont
  {{Glazman}}},\ }\href {\doibase 10.1103/PhysRevLett.110.216402} {\bibfield
  {journal} {\bibinfo  {journal} {Physical Review Letters}\ }\textbf {\bibinfo
  {volume} {110}},\ \bibinfo {eid} {216402} (\bibinfo {year} {2013})},\ \Eprint
  {http://arxiv.org/abs/1303.1766} {arXiv:1303.1766} \BibitemShut {NoStop}%
\bibitem [{\citenamefont {{Altshuler}}\ \emph {et~al.}(2013)\citenamefont
  {{Altshuler}}, \citenamefont {{Aleiner}},\ and\ \citenamefont
  {{Yudson}}}]{2013PhRvL.111h6401A}%
  \BibitemOpen
  \bibfield  {author} {\bibinfo {author} {\bibfnamefont {B.~L.}\ \bibnamefont
  {{Altshuler}}}, \bibinfo {author} {\bibfnamefont {I.~L.}\ \bibnamefont
  {{Aleiner}}}, \ and\ \bibinfo {author} {\bibfnamefont {V.~I.}\ \bibnamefont
  {{Yudson}}},\ }\href {\doibase 10.1103/PhysRevLett.111.086401} {\bibfield
  {journal} {\bibinfo  {journal} {Physical Review Letters}\ }\textbf {\bibinfo
  {volume} {111}},\ \bibinfo {eid} {086401} (\bibinfo {year} {2013})},\ \Eprint
  {http://arxiv.org/abs/1306.2626} {arXiv:1306.2626} \BibitemShut {NoStop}%
\bibitem [{\citenamefont {{V{\"a}yrynen}}\ \emph {et~al.}(2014)\citenamefont
  {{V{\"a}yrynen}}, \citenamefont {{Goldstein}}, \citenamefont {{Gefen}},\ and\
  \citenamefont {{Glazman}}}]{2014PhRvB..90k5309V}%
  \BibitemOpen
  \bibfield  {author} {\bibinfo {author} {\bibfnamefont {J.~I.}\ \bibnamefont
  {{V{\"a}yrynen}}}, \bibinfo {author} {\bibfnamefont {M.}~\bibnamefont
  {{Goldstein}}}, \bibinfo {author} {\bibfnamefont {Y.}~\bibnamefont
  {{Gefen}}}, \ and\ \bibinfo {author} {\bibfnamefont {L.~I.}\ \bibnamefont
  {{Glazman}}},\ }\href {\doibase 10.1103/PhysRevB.90.115309} {\bibfield
  {journal} {\bibinfo  {journal} {\prb}\ }\textbf {\bibinfo {volume} {90}},\
  \bibinfo {eid} {115309} (\bibinfo {year} {2014})},\ \Eprint
  {http://arxiv.org/abs/1406.6052} {arXiv:1406.6052} \BibitemShut {NoStop}%
\bibitem [{\citenamefont {{Kainaris}}\ \emph {et~al.}(2014)\citenamefont
  {{Kainaris}}, \citenamefont {{Gornyi}}, \citenamefont {{Carr}},\ and\
  \citenamefont {{Mirlin}}}]{2014PhRvB..90g5118K}%
  \BibitemOpen
  \bibfield  {author} {\bibinfo {author} {\bibfnamefont {N.}~\bibnamefont
  {{Kainaris}}}, \bibinfo {author} {\bibfnamefont {I.~V.}\ \bibnamefont
  {{Gornyi}}}, \bibinfo {author} {\bibfnamefont {S.~T.}\ \bibnamefont
  {{Carr}}}, \ and\ \bibinfo {author} {\bibfnamefont {A.~D.}\ \bibnamefont
  {{Mirlin}}},\ }\href {\doibase 10.1103/PhysRevB.90.075118} {\bibfield
  {journal} {\bibinfo  {journal} {\prb}\ }\textbf {\bibinfo {volume} {90}},\
  \bibinfo {eid} {075118} (\bibinfo {year} {2014})},\ \Eprint
  {http://arxiv.org/abs/1404.3129} {arXiv:1404.3129} \BibitemShut {NoStop}%
\bibitem [{\citenamefont {{Chou}}\ \emph {et~al.}(2015)\citenamefont {{Chou}},
  \citenamefont {{Levchenko}},\ and\ \citenamefont
  {{Foster}}}]{2015PhRvL.115r6404C}%
  \BibitemOpen
  \bibfield  {author} {\bibinfo {author} {\bibfnamefont {Y.-Z.}\ \bibnamefont
  {{Chou}}}, \bibinfo {author} {\bibfnamefont {A.}~\bibnamefont {{Levchenko}}},
  \ and\ \bibinfo {author} {\bibfnamefont {M.~S.}\ \bibnamefont {{Foster}}},\
  }\href {\doibase 10.1103/PhysRevLett.115.186404} {\bibfield  {journal}
  {\bibinfo  {journal} {Physical Review Letters}\ }\textbf {\bibinfo {volume}
  {115}},\ \bibinfo {eid} {186404} (\bibinfo {year} {2015})},\ \Eprint
  {http://arxiv.org/abs/1508.04433} {arXiv:1508.04433 [cond-mat.mes-hall]}
  \BibitemShut {NoStop}%
\bibitem [{Note1()}]{Note1}%
  \BibitemOpen
  \bibinfo {note} {The spin quantization can also be broken in a disordered
  manner; in this paper we ignore these terms for simplicity. Note that they do
  not lead to significant backscattering at low temperature~\cite
  {PhysRevLett.116.086603}. Electron-electron interactions together with Rashba
  spin-orbit disorder lead to a correction to conductance that vanishes as
  $T^4$ (or with a higher power law) as $T \to 0$~\cite
  {PhysRevB.96.155134}.}\BibitemShut {Stop}%
\bibitem [{\citenamefont {Lezmy}\ \emph {et~al.}(2012)\citenamefont {Lezmy},
  \citenamefont {Oreg},\ and\ \citenamefont {Berkooz}}]{PhysRevB.85.235304}%
  \BibitemOpen
  \bibfield  {author} {\bibinfo {author} {\bibfnamefont {N.}~\bibnamefont
  {Lezmy}}, \bibinfo {author} {\bibfnamefont {Y.}~\bibnamefont {Oreg}}, \ and\
  \bibinfo {author} {\bibfnamefont {M.}~\bibnamefont {Berkooz}},\ }\href
  {\doibase 10.1103/PhysRevB.85.235304} {\bibfield  {journal} {\bibinfo
  {journal} {Phys. Rev. B}\ }\textbf {\bibinfo {volume} {85}},\ \bibinfo
  {pages} {235304} (\bibinfo {year} {2012})}\BibitemShut {NoStop}%
\bibitem [{Note2()}]{Note2}%
  \BibitemOpen
  \bibinfo {note} {Localized spins may occur for example due to charge
  puddles~\cite {2013PhRvL.110u6402V,2014PhRvB..90k5309V} or nuclear
  spins~\cite {Hsu2017,*Hsu2018}}\BibitemShut {NoStop}%
\bibitem [{\citenamefont {Xie}\ \emph {et~al.}(2016)\citenamefont {Xie},
  \citenamefont {Li}, \citenamefont {Chou},\ and\ \citenamefont
  {Foster}}]{PhysRevLett.116.086603}%
  \BibitemOpen
  \bibfield  {author} {\bibinfo {author} {\bibfnamefont {H.-Y.}\ \bibnamefont
  {Xie}}, \bibinfo {author} {\bibfnamefont {H.}~\bibnamefont {Li}}, \bibinfo
  {author} {\bibfnamefont {Y.-Z.}\ \bibnamefont {Chou}}, \ and\ \bibinfo
  {author} {\bibfnamefont {M.~S.}\ \bibnamefont {Foster}},\ }\href {\doibase
  10.1103/PhysRevLett.116.086603} {\bibfield  {journal} {\bibinfo  {journal}
  {Phys. Rev. Lett.}\ }\textbf {\bibinfo {volume} {116}},\ \bibinfo {pages}
  {086603} (\bibinfo {year} {2016})}\BibitemShut {NoStop}%
\bibitem [{\citenamefont {Kubo}(1957)}]{kubo1957statistical}%
  \BibitemOpen
  \bibfield  {author} {\bibinfo {author} {\bibfnamefont {R.}~\bibnamefont
  {Kubo}},\ }\href@noop {} {\bibfield  {journal} {\bibinfo  {journal} {Journal
  of the Physical Society of Japan}\ }\textbf {\bibinfo {volume} {12}},\
  \bibinfo {pages} {570} (\bibinfo {year} {1957})}\BibitemShut {NoStop}%
\bibitem [{Note3()}]{Note3}%
  \BibitemOpen
  \bibinfo {note} {See Supplementary Material, where we consider noise from
  fluctuating dipoles in the dielectric as well as a quantum mechanical
  treatment of $w(t)$.}\BibitemShut {Stop}%
\bibitem [{\citenamefont {Kogan}(1996)}]{9780511551666}%
  \BibitemOpen
  \bibfield  {author} {\bibinfo {author} {\bibfnamefont {S.}~\bibnamefont
  {Kogan}},\ }\href {http://dx.doi.org/10.1017/CBO9780511551666} {\emph
  {\bibinfo {title} {Electronic Noise and Fluctuations in Solids}}}\ (\bibinfo
  {publisher} {Cambridge University Press},\ \bibinfo {year}
  {1996})\BibitemShut {NoStop}%
\bibitem [{\citenamefont {Li}\ \emph {et~al.}(2015)\citenamefont {Li},
  \citenamefont {Wang}, \citenamefont {Fu}, \citenamefont {Du}, \citenamefont
  {Schreiber}, \citenamefont {Mu}, \citenamefont {Liu}, \citenamefont
  {Sullivan}, \citenamefont {Cs\'athy}, \citenamefont {Lin},\ and\
  \citenamefont {Du}}]{Li15}%
  \BibitemOpen
  \bibfield  {author} {\bibinfo {author} {\bibfnamefont {T.}~\bibnamefont
  {Li}}, \bibinfo {author} {\bibfnamefont {P.}~\bibnamefont {Wang}}, \bibinfo
  {author} {\bibfnamefont {H.}~\bibnamefont {Fu}}, \bibinfo {author}
  {\bibfnamefont {L.}~\bibnamefont {Du}}, \bibinfo {author} {\bibfnamefont
  {K.~A.}\ \bibnamefont {Schreiber}}, \bibinfo {author} {\bibfnamefont
  {X.}~\bibnamefont {Mu}}, \bibinfo {author} {\bibfnamefont {X.}~\bibnamefont
  {Liu}}, \bibinfo {author} {\bibfnamefont {G.}~\bibnamefont {Sullivan}},
  \bibinfo {author} {\bibfnamefont {G.~A.}\ \bibnamefont {Cs\'athy}}, \bibinfo
  {author} {\bibfnamefont {X.}~\bibnamefont {Lin}}, \ and\ \bibinfo {author}
  {\bibfnamefont {R.-R.}\ \bibnamefont {Du}},\ }\href {\doibase
  10.1103/PhysRevLett.115.136804} {\bibfield  {journal} {\bibinfo  {journal}
  {Phys. Rev. Lett.}\ }\textbf {\bibinfo {volume} {115}},\ \bibinfo {pages}
  {136804} (\bibinfo {year} {2015})}\BibitemShut {NoStop}%
\bibitem [{\citenamefont {Toda}\ \emph {et~al.}(2012)\citenamefont {Toda},
  \citenamefont {Kubo}, \citenamefont {Kubo}, \citenamefont {Toda},
  \citenamefont {Saito}, \citenamefont {Hashitsume},\ and\ \citenamefont
  {Hashitsume}}]{toda2012statistical}%
  \BibitemOpen
  \bibfield  {author} {\bibinfo {author} {\bibfnamefont {M.}~\bibnamefont
  {Toda}}, \bibinfo {author} {\bibfnamefont {R.}~\bibnamefont {Kubo}}, \bibinfo
  {author} {\bibfnamefont {R.}~\bibnamefont {Kubo}}, \bibinfo {author}
  {\bibfnamefont {M.}~\bibnamefont {Toda}}, \bibinfo {author} {\bibfnamefont
  {N.}~\bibnamefont {Saito}}, \bibinfo {author} {\bibfnamefont
  {N.}~\bibnamefont {Hashitsume}}, \ and\ \bibinfo {author} {\bibfnamefont
  {N.}~\bibnamefont {Hashitsume}},\ }\href@noop {} {\emph {\bibinfo {title}
  {Statistical Physics II: Nonequilibrium Statistical Mechanics}}},\ Springer
  Series in Solid-State Sciences\ (\bibinfo  {publisher} {Springer Berlin
  Heidelberg},\ \bibinfo {year} {2012})\BibitemShut {NoStop}%
\bibitem [{\citenamefont {Forster}(1995)}]{forster2018hydrodynamic}%
  \BibitemOpen
  \bibfield  {author} {\bibinfo {author} {\bibfnamefont {D.}~\bibnamefont
  {Forster}},\ }\href@noop {} {\emph {\bibinfo {title} {Hydrodynamic
  Fluctuations, Broken Symmetry, And Correlation Functions}}}\ (\bibinfo
  {publisher} {CRC Press},\ \bibinfo {year} {1995})\BibitemShut {NoStop}%
\bibitem [{Note4()}]{Note4}%
  \BibitemOpen
  \bibinfo {note} {For example, in the case of two TLSs, of the form Eq.~(\ref
  {eq:Hvwithtexture}), the interference term will be proportional to
  $w_{1}(t)w_{2}(t+t')$, whose average over $t$ is independent of $t'$ and
  therefore it doesn't contribute to $\protect \overline {\delimiter "426830A
  \delta I\delimiter "526930B }$ for the same reason as a static perturbation
  does not contribute to it.}\BibitemShut {Stop}%
\bibitem [{Note5()}]{Note5}%
  \BibitemOpen
  \bibinfo {note} {The density $n$ is obtained from the two-dimensional density
  $n_{2D}$ by multiplying the latter with the effective range $\xi $ of the
  potential $U$; $n=n_{2D}\protect \tmspace +\thinmuskip {.1667em}\xi $. In
  practice, $\xi $ is the screening length given by the distance to nearest
  gate.}\BibitemShut {Stop}%
\bibitem [{\citenamefont {Rod}\ \emph {et~al.}(2015)\citenamefont {Rod},
  \citenamefont {Schmidt},\ and\ \citenamefont {Rachel}}]{PhysRevB.91.245112}%
  \BibitemOpen
  \bibfield  {author} {\bibinfo {author} {\bibfnamefont {A.}~\bibnamefont
  {Rod}}, \bibinfo {author} {\bibfnamefont {T.~L.}\ \bibnamefont {Schmidt}}, \
  and\ \bibinfo {author} {\bibfnamefont {S.}~\bibnamefont {Rachel}},\ }\href
  {\doibase 10.1103/PhysRevB.91.245112} {\bibfield  {journal} {\bibinfo
  {journal} {Phys. Rev. B}\ }\textbf {\bibinfo {volume} {91}},\ \bibinfo
  {pages} {245112} (\bibinfo {year} {2015})}\BibitemShut {NoStop}%
\bibitem [{\citenamefont {{Skolasinski}}\ \emph {et~al.}(2017)\citenamefont
  {{Skolasinski}}, \citenamefont {{Pikulin}}, \citenamefont {{Alicea}},\ and\
  \citenamefont {{Wimmer}}}]{2017arXiv170904830S}%
  \BibitemOpen
  \bibfield  {author} {\bibinfo {author} {\bibfnamefont {R.}~\bibnamefont
  {{Skolasinski}}}, \bibinfo {author} {\bibfnamefont {D.~I.}\ \bibnamefont
  {{Pikulin}}}, \bibinfo {author} {\bibfnamefont {J.}~\bibnamefont {{Alicea}}},
  \ and\ \bibinfo {author} {\bibfnamefont {M.}~\bibnamefont {{Wimmer}}},\
  }\href@noop {} {\bibfield  {journal} {\bibinfo  {journal} {ArXiv e-prints}\ }
  (\bibinfo {year} {2017})},\ \Eprint {http://arxiv.org/abs/1709.04830}
  {arXiv:1709.04830} \BibitemShut {NoStop}%
\bibitem [{Note6()}]{Note6}%
  \BibitemOpen
  \bibinfo {note} {From Ref.~\cite {2014PhRvB..90k5309V} we have $\delta
  \approx \alpha ^2 \Delta _{b}$ and $\alpha \approx 0.3$.}\BibitemShut {Stop}%
\bibitem [{Note7()}]{Note7}%
  \BibitemOpen
  \bibinfo {note} {For TLS in the bulk of the 2D TI, one can set $d_0 \to 0$
  for TLS that are very close to the edge.}\BibitemShut {Stop}%
\bibitem [{\citenamefont {Stephens}(1973)}]{PhysRevB.8.2896}%
  \BibitemOpen
  \bibfield  {author} {\bibinfo {author} {\bibfnamefont {R.~B.}\ \bibnamefont
  {Stephens}},\ }\href {\doibase 10.1103/PhysRevB.8.2896} {\bibfield  {journal}
  {\bibinfo  {journal} {Phys. Rev. B}\ }\textbf {\bibinfo {volume} {8}},\
  \bibinfo {pages} {2896} (\bibinfo {year} {1973})}\BibitemShut {NoStop}%
\bibitem [{\citenamefont {Constantin}\ \emph {et~al.}(2009)\citenamefont
  {Constantin}, \citenamefont {Yu},\ and\ \citenamefont
  {Martinis}}]{PhysRevB.79.094520}%
  \BibitemOpen
  \bibfield  {author} {\bibinfo {author} {\bibfnamefont {M.}~\bibnamefont
  {Constantin}}, \bibinfo {author} {\bibfnamefont {C.~C.}\ \bibnamefont {Yu}},
  \ and\ \bibinfo {author} {\bibfnamefont {J.~M.}\ \bibnamefont {Martinis}},\
  }\href {\doibase 10.1103/PhysRevB.79.094520} {\bibfield  {journal} {\bibinfo
  {journal} {Phys. Rev. B}\ }\textbf {\bibinfo {volume} {79}},\ \bibinfo
  {pages} {094520} (\bibinfo {year} {2009})}\BibitemShut {NoStop}%
\bibitem [{\citenamefont {Garcia}\ \emph {et~al.}(2018)\citenamefont {Garcia},
  \citenamefont {Vila}, \citenamefont {Cummings},\ and\ \citenamefont
  {Roche}}]{garcia2018spin}%
  \BibitemOpen
  \bibfield  {author} {\bibinfo {author} {\bibfnamefont {J.~H.}\ \bibnamefont
  {Garcia}}, \bibinfo {author} {\bibfnamefont {M.}~\bibnamefont {Vila}},
  \bibinfo {author} {\bibfnamefont {A.~W.}\ \bibnamefont {Cummings}}, \ and\
  \bibinfo {author} {\bibfnamefont {S.}~\bibnamefont {Roche}},\ }\href@noop {}
  {\bibfield  {journal} {\bibinfo  {journal} {Chemical Society Reviews}\
  }\textbf {\bibinfo {volume} {47}},\ \bibinfo {pages} {3359} (\bibinfo {year}
  {2018})}\BibitemShut {NoStop}%
\bibitem [{\citenamefont {Zhang}\ \emph {et~al.}(2009)\citenamefont {Zhang},
  \citenamefont {Cheng}, \citenamefont {Chen}, \citenamefont {Jia},
  \citenamefont {Ma}, \citenamefont {He}, \citenamefont {Wang}, \citenamefont
  {Zhang}, \citenamefont {Dai}, \citenamefont {Fang}, \citenamefont {Xie},\
  and\ \citenamefont {Xue}}]{PhysRevLett.103.266803}%
  \BibitemOpen
  \bibfield  {author} {\bibinfo {author} {\bibfnamefont {T.}~\bibnamefont
  {Zhang}}, \bibinfo {author} {\bibfnamefont {P.}~\bibnamefont {Cheng}},
  \bibinfo {author} {\bibfnamefont {X.}~\bibnamefont {Chen}}, \bibinfo {author}
  {\bibfnamefont {J.-F.}\ \bibnamefont {Jia}}, \bibinfo {author} {\bibfnamefont
  {X.}~\bibnamefont {Ma}}, \bibinfo {author} {\bibfnamefont {K.}~\bibnamefont
  {He}}, \bibinfo {author} {\bibfnamefont {L.}~\bibnamefont {Wang}}, \bibinfo
  {author} {\bibfnamefont {H.}~\bibnamefont {Zhang}}, \bibinfo {author}
  {\bibfnamefont {X.}~\bibnamefont {Dai}}, \bibinfo {author} {\bibfnamefont
  {Z.}~\bibnamefont {Fang}}, \bibinfo {author} {\bibfnamefont {X.}~\bibnamefont
  {Xie}}, \ and\ \bibinfo {author} {\bibfnamefont {Q.-K.}\ \bibnamefont
  {Xue}},\ }\href {\doibase 10.1103/PhysRevLett.103.266803} {\bibfield
  {journal} {\bibinfo  {journal} {Phys. Rev. Lett.}\ }\textbf {\bibinfo
  {volume} {103}},\ \bibinfo {pages} {266803} (\bibinfo {year}
  {2009})}\BibitemShut {NoStop}%
\bibitem [{\citenamefont {Alpichshev}\ \emph {et~al.}(2010)\citenamefont
  {Alpichshev}, \citenamefont {Analytis}, \citenamefont {Chu}, \citenamefont
  {Fisher}, \citenamefont {Chen}, \citenamefont {Shen}, \citenamefont {Fang},\
  and\ \citenamefont {Kapitulnik}}]{PhysRevLett.104.016401}%
  \BibitemOpen
  \bibfield  {author} {\bibinfo {author} {\bibfnamefont {Z.}~\bibnamefont
  {Alpichshev}}, \bibinfo {author} {\bibfnamefont {J.~G.}\ \bibnamefont
  {Analytis}}, \bibinfo {author} {\bibfnamefont {J.-H.}\ \bibnamefont {Chu}},
  \bibinfo {author} {\bibfnamefont {I.~R.}\ \bibnamefont {Fisher}}, \bibinfo
  {author} {\bibfnamefont {Y.~L.}\ \bibnamefont {Chen}}, \bibinfo {author}
  {\bibfnamefont {Z.~X.}\ \bibnamefont {Shen}}, \bibinfo {author}
  {\bibfnamefont {A.}~\bibnamefont {Fang}}, \ and\ \bibinfo {author}
  {\bibfnamefont {A.}~\bibnamefont {Kapitulnik}},\ }\href {\doibase
  10.1103/PhysRevLett.104.016401} {\bibfield  {journal} {\bibinfo  {journal}
  {Phys. Rev. Lett.}\ }\textbf {\bibinfo {volume} {104}},\ \bibinfo {pages}
  {016401} (\bibinfo {year} {2010})}\BibitemShut {NoStop}%
\bibitem [{\citenamefont {Kharitonov}\ \emph {et~al.}(2017)\citenamefont
  {Kharitonov}, \citenamefont {Geissler},\ and\ \citenamefont
  {Trauzettel}}]{PhysRevB.96.155134}%
  \BibitemOpen
  \bibfield  {author} {\bibinfo {author} {\bibfnamefont {M.}~\bibnamefont
  {Kharitonov}}, \bibinfo {author} {\bibfnamefont {F.}~\bibnamefont
  {Geissler}}, \ and\ \bibinfo {author} {\bibfnamefont {B.}~\bibnamefont
  {Trauzettel}},\ }\href {\doibase 10.1103/PhysRevB.96.155134} {\bibfield
  {journal} {\bibinfo  {journal} {Phys. Rev. B}\ }\textbf {\bibinfo {volume}
  {96}},\ \bibinfo {pages} {155134} (\bibinfo {year} {2017})}\BibitemShut
  {NoStop}%
\bibitem [{\citenamefont {Hsu}\ \emph {et~al.}(2017)\citenamefont {Hsu},
  \citenamefont {Stano}, \citenamefont {Klinovaja},\ and\ \citenamefont
  {Loss}}]{Hsu2017}%
  \BibitemOpen
  \bibfield  {author} {\bibinfo {author} {\bibfnamefont {C.-H.}\ \bibnamefont
  {Hsu}}, \bibinfo {author} {\bibfnamefont {P.}~\bibnamefont {Stano}}, \bibinfo
  {author} {\bibfnamefont {J.}~\bibnamefont {Klinovaja}}, \ and\ \bibinfo
  {author} {\bibfnamefont {D.}~\bibnamefont {Loss}},\ }\href {\doibase
  10.1103/PhysRevB.96.081405} {\bibfield  {journal} {\bibinfo  {journal} {Phys.
  Rev. B}\ }\textbf {\bibinfo {volume} {96}},\ \bibinfo {pages} {081405}
  (\bibinfo {year} {2017})}\BibitemShut {NoStop}%
\bibitem [{\citenamefont {Hsu}\ \emph {et~al.}(2018)\citenamefont {Hsu},
  \citenamefont {Stano}, \citenamefont {Klinovaja},\ and\ \citenamefont
  {Loss}}]{Hsu2018}%
  \BibitemOpen
  \bibfield  {author} {\bibinfo {author} {\bibfnamefont {C.-H.}\ \bibnamefont
  {Hsu}}, \bibinfo {author} {\bibfnamefont {P.}~\bibnamefont {Stano}}, \bibinfo
  {author} {\bibfnamefont {J.}~\bibnamefont {Klinovaja}}, \ and\ \bibinfo
  {author} {\bibfnamefont {D.}~\bibnamefont {Loss}},\ }\href {\doibase
  10.1103/PhysRevB.97.125432} {\bibfield  {journal} {\bibinfo  {journal} {Phys.
  Rev. B}\ }\textbf {\bibinfo {volume} {97}},\ \bibinfo {pages} {125432}
  (\bibinfo {year} {2018})}\BibitemShut {NoStop}%
\end{thebibliography}%



\section*{Supplementary material}

# Supplemental Material to "Noise-induced backscattering in a quantum-spin-Hall edge"

Jukka I. Väyrynen and Dmitry I. Pikulin
*Station Q, Microsoft Research, Santa Barbara, California 93106-6105, USA*

Jason Alicea
*Department of Physics and Institute for Quantum Information and Matter,
California Institute of Technology, Pasadena, CA 91125, USA and
Walter Burke Institute for Theoretical Physics, California Institute of Technology, Pasadena, CA 91125, USA*

In this supplemental material we show in detail the derivations that were omitted in the main text.

## SM1. POTENTIAL AND CONDUCTANCE CORRECTION FROM A SINGLE DIPOLE IN THE DIELECTRIC

Let us start from the Hamiltonian

$$H_U(t) = \int dx \rho(x) U(x) w(t) , \tag{S1}$$

where now the potential $U(x)w(t)$ results from a single fluctuating dipole. We have then

$$U(x)w(t) = \frac{e}{4\pi\varepsilon} \frac{\mathbf{p}_i(t) \cdot (\mathbf{x} - \mathbf{r}_i)}{|\mathbf{x} - \mathbf{r}_i|^3} . \tag{S2}$$

Here $\mathbf{p}_i, \mathbf{r}_i$ are the dipole moment and position of the $i$th dipole and $\varepsilon$ is the dielectric constant of the dielectric. We will assume that the dipoles fluctuate independently,

$$\overline{\mathbf{p}_i(t)\mathbf{p}_j(t+t')} = \delta_{ij} \mathbf{p}_i \mathbf{p}_i \overline{w(t)w(t+t')} . \tag{S3}$$

This assumption allows us to calculate first the single-dipole contribution and then sum those contributions to get the edge resistance. With the above form we can use Eq. (9) of the main text generalized to $\omega \gtrsim T$ [see Eq. (S22) and below],

$$\delta G = \frac{G_0}{v^2} |U_{2k_F}|^2 n_\perp^2 D^{-2} \int \frac{d\omega}{2\pi} \omega^2 \frac{(\omega/2T)^2}{\sinh^2(\omega/2T)} S(\omega) , \tag{S4}$$

where $\int dx U(x) e^{-i2k_F x} = U_{2k_F}$ and

$$S(\omega) = \int_{-\infty}^{\infty} dt' e^{i\omega t'} \overline{w(t)w(t'+t)} \tag{S5}$$

is the classical noise spectrum.

Let us calculate $|U_{2k_F}|^2$. We have

$$U_{2k_F} = \frac{e}{4\pi\varepsilon} \int dx \frac{\mathbf{p}_i \cdot (\mathbf{x} - \mathbf{r}_i)}{|\mathbf{x} - \mathbf{r}_i|^3} e^{-i2k_F x} . \tag{S6}$$

Let's shift $x \to x + r_{i,x}$. Denoting $\rho_i^2 = r_{i,z}^2 + r_{i,y}^2$ and $\boldsymbol{\rho} = (r_{i,y}, r_{i,z})$, we find the $x$-integral in terms of Bessel functions,

$$U_{2k_F} = \frac{e}{4\pi\varepsilon} e^{i2k_F r_{i,x}} \int dx e^{-i2k_F x} \frac{(p_{i,x} x - \mathbf{p}_i \cdot \boldsymbol{\rho})}{\sqrt{x^2 + \rho_i^2}^3} \tag{S7}$$

$$= 4k_F \frac{e}{4\pi\varepsilon} e^{i2k_F r_{i,x}} \left( p_{i,x} i K_0 (2k_F |\rho_i|) - \frac{\mathbf{p}_i \cdot \boldsymbol{\rho}_i}{|\boldsymbol{\rho}_i|} K_1 (2k_F |\rho_i|) \right) . \tag{S8}$$

The phase $e^{i2k_F r_{i,x}}$ cancels out when we take absolute value squared,

$$|U_{2k_F}|^2 = 16k_F^2 \left(\frac{e}{4\pi\varepsilon}\right)^2 \left( p_{i,x}^2 K_0 (2k_F|\rho_i|)^2 + \left(\frac{\mathbf{p}_i \cdot \boldsymbol{\rho}_i}{|\boldsymbol{\rho}_i|}\right)^2 K_1 (2k_F|\rho_i|)^2 \right) . \tag{S9}$$

## SM2. RESISTANCE FROM A COLLECTION OF DIPOLES

In the presence of many dipoles, we obtain a resistance as a sum over dipoles,

$$R = \sum_i \frac{\delta G_i}{G_0^2}. \tag{S10}$$

Introducing dipole density $n(\mathbf{r}) = \sum_i \delta(\mathbf{r}_i - \mathbf{r})$, we find

$$R = \frac{1}{G_0} \frac{1}{D^2 v^2} 16 k_F^2 \left(\frac{e}{4\pi\varepsilon}\right)^2 \frac{1}{3} L n p^2 \int d^2\boldsymbol{\rho} \left( K_0 \left(2k_F|\rho|\right)^2 + 2K_1 \left(2k_F|\rho|\right)^2 \right) n_\perp^2 \int \frac{d\omega}{2\pi} \omega^2 \frac{(\omega/2T)^2}{\sinh^2(\omega/2T)} S(\omega), \tag{S11}$$

where we assumed a uniform density $n(\mathbf{r}) = n$ in a volume $\mathbf{r} \in (0,L) \times (d, d+H) \times (0, W)$. We assume that the dielectric is of length $L$, width $W$, thickness $H$, and separated by a distance $d$ from the helical edge. We assumed equal strengths $p$ and random directions for the dipole moments, $p_a p_b \to \delta_{ab} \frac{1}{3} p^2$.

Assuming $H, W \gg k_F^{-1}$, we can set $\int d^2\boldsymbol{\rho} = \frac{\pi}{2} \int_d^\infty d\rho \rho$. Then the integral over $\rho$ can be done in the limit of large $k_F d$,

$$\int d^2\boldsymbol{\rho} \left( K_0 \left(2k_F|\rho|\right)^2 + 2K_1 \left(2k_F|\rho|\right)^2 \right) \approx \frac{1}{4k_F^2} \frac{\pi^2}{8} e^{-4k_F d}. \tag{S12}$$

The resistance becomes

$$R = \frac{1}{G_0} \frac{1}{D^2 v^2} 4 \left(\frac{e}{4\pi\varepsilon}\right)^2 \frac{1}{3} L n \frac{\pi^2}{8} e^{-4dk_F} p^2 n_\perp^2 \int \frac{d\omega}{2\pi} \omega^2 \frac{(\omega/2T)^2}{\sinh^2(\omega/2T)} S(\omega). \tag{S13}$$

From Ref. [S1], we find

$$p^2 S(\omega) = V^2 \frac{1}{2} \frac{1}{|\omega|} S_0, \quad S_0 = \frac{4\varepsilon T}{V} \tan\delta, \tag{S14}$$

where $\tan\delta$ is the dielectric loss tangent and $V = WHL$ the volume.

Integration over $\omega$ yields (denoting $N = nV$ the total number of contributing dipoles)

$$R = \frac{\zeta(3)}{2} \frac{1}{G_0} \frac{L}{v/T} \frac{e^2}{4\pi\varepsilon v} N e^{-4dk_F} n_\perp^2 \frac{T^2}{D^2} \tan\delta. \tag{S15}$$

where $\zeta$ is the Riemann zeta function, $\zeta(3) \approx 1.2$. This equation is used in the discussion section of the main text.

## SM3. GENERALIZATION OF EQ. (9) TO FREQUENCIES $\omega \gtrsim T$

Eq. (7) of the main text,

$$\langle \delta I(t) \rangle = e \int \frac{dk}{2\pi} \frac{dk'}{2\pi} |[B_{k'}^\dagger B_k]_{10}|^2 |U_{k-k'}|^2 (f_{kL} - f_{-kR}) 2\text{Re} \int_{-\infty}^0 dt' e^{i(vk+vk'-i0)t'} w(t) w(t'+t). \tag{S16}$$

derived from

$$H_U(t) = \int dx \rho(x) U(x) w(t), \tag{S17}$$

can be generalized to the case where $\hat{w}$ is an operator. Assuming $\hat{w}(t)$ and the Hamiltonian governing its dynamics does not couple to edge electrons, we find

$$\langle \delta I(t) \rangle = e \int \frac{dk}{2\pi} \frac{dk'}{2\pi} |[B_{k'}^\dagger B_k]_{10}|^2 |U_{k-k'}|^2 \sum_\alpha \alpha f_{-k'\overline{\alpha}} [1 - f_{-k\alpha}] S_Q(\alpha v(k+k')), \tag{S18}$$

where

$$S_Q(\omega) = 2\text{Re} \int_{-\infty}^0 dt' \langle \hat{w}(t'+t) \hat{w}(t) \rangle e^{i(\omega - i0)t'}. \tag{S19}$$

This spectral function has the property ($\beta = 1/T$)

$$S_Q(\omega) = e^{\beta\omega} S_Q(-\omega). \tag{S20}$$

By using this property and the assumptions made when deriving Eq. (9) of the main text, we can derive for backscattering current,

$$\langle \delta I \rangle = \frac{e}{2\pi} \left( e^{\beta eV} - 1 \right) v^{-2} n_\perp^2 D^{-2} |U_{2k_F}|^2 \int \frac{d\omega}{2\pi} \omega^2 \frac{e^{-\beta(\omega+eV)}(\omega+eV)}{e^{-\beta(\omega+eV)} - 1} S_Q(\omega), \tag{S21}$$

which in the linear response limit becomes

$$\langle \delta I \rangle = G_0 V v^{-2} n_\perp^2 D^{-2} |U_{2k_F}|^2 \int \frac{d\omega}{2\pi} \omega^2 \frac{e^{-\beta\omega} \beta\omega}{e^{-\beta\omega} - 1} S_Q(\omega). \tag{S22}$$

For a quantum two-level system, we have [S2]

$$S_Q(\omega) = \frac{-\beta\omega}{e^{-\beta\omega} - 1} S(\omega), \tag{S23}$$

where $S(\omega) = p_0(1-p_0)\frac{2\tau}{1+\omega^2\tau^2}$ is the TLS noise spectrum. In the classical limit $\beta\omega \to 0$, we have $S_Q(\omega) = S(\omega)$. Finally, using the identity

$$\frac{\beta\omega}{e^{-\beta\omega} - 1} \frac{e^{-\beta\omega} \beta\omega}{e^{-\beta\omega} - 1} = \frac{(\omega/2T)^2}{\sinh^2(\omega/2T)}, \tag{S24}$$

we obtain the replacement given in the main text.

---

[S1] M. Constantin, C. C. Yu, and J. M. Martinis, Phys. Rev. B **79**, 094520 (2009).
[S2] D. Forster, *Hydrodynamic Fluctuations, Broken Symmetry, And Correlation Functions* (CRC Press, 1995).



\end{document}